\documentclass[11pt]{article}
\usepackage{geometry} % see geometry.pdf on how to lay out the page. There's lots.
\usepackage[english]{babel}
\usepackage[english]{layout}
\usepackage{cite}

\usepackage{color}
\usepackage{mathpazo} % math & rm
\usepackage[scaled]{helvet} % ss
\usepackage{courier} % tt
\normalfont
\usepackage[T1]{fontenc}

\usepackage{amsfonts,latexsym,amsthm,amssymb}
\usepackage{amsmath,bbm}  % Better maths support & more symbols

\usepackage{memhfixc}  % remove conflict between the memoir class & hyperref
\usepackage{fancyhdr} % This should be set AFTER setting up the page geometry
\usepackage{sectsty}
\allsectionsfont{\sffamily\mdseries\upshape} % (See the fntguide.pdf for font help)
\numberwithin{equation}{section}

\begin{document}

\title{\LARGE \textbf{Algebraic Bethe ansatz for the trigonometric $s\ell(2)$ Gaudin model with triangular boundary}}

\author{
%\textsf{N. ~Cirilo ~Ant\'onio,}
%\thanks{E-mail address: nantonio@math.ist.utl.pt}
\textsf{ ~ N. ~Manojlovi\'c,}
\thanks{E-mail address: nmanoj@ualg.pt}
%\textsf{ ~ E. ~Ragoucy}
%\thanks{E-mail address: eric.ragoucy@lapth.cnrs.fr}
\textsf{~ and I. ~Salom}
\thanks{E-mail address: isalom@ipb.ac.rs} \\
%\\
%\textit{$^{\ast}$Centro de An\'alise Funcional e Aplica\c{c}\~oes}\\
%\textit{Instituto Superior T\'ecnico, Universidade de Lisboa} \\
%\textit{Av. Rovisco Pais, 1049-001 Lisboa, Portugal} \\
%\\
%\textit{$^{\dag}$Department of Mathematics, Massachusetts Institute of Technology} \\
%\textit{77 Massachusetts Avenue, Cambridge, MA 02139-4307, USA.} \\
%\\
%\textit{$^{\dag}$Grupo de F\'{\i}sica Matem\'atica da Universidade de Lisboa} \\
%\textit{Av. Prof. Gama Pinto 2, PT-1649-003 Lisboa, Portugal} \\
\\
\textit{$^{\dag}$Departamento de Matem\'atica, F. C. T.,
Universidade do Algarve} \\
\textit{Campus de Gambelas, PT-8005-139 Faro, Portugal}\\
%\\
%\textit{$^{\ddag}$Laboratoire d'Annecy-le-Vieux de Physique Th\'eorique LAPTh}\\
%\textit{CNRS et Univerit\'e de Savoie, UMR 5108, B.P. 110}\\ 
%\textit{74941 Annecy-le-Vieux Cedex, France}\\
\\
\textit{$^{\S}$Institute of Physics, University of Belgrade}\\
\textit{P.O. Box 57, 11080 Belgrade, Serbia}\\
\\
}

\date{}

\maketitle
\thispagestyle{empty}

\begin{abstract}
In the derivation of the generating function of the Gaudin Hamiltonians with boundary terms we follow the same approach used previously in the rational case, which in turn was based on Sklyanin's method in the periodic case. Our derivation is centered on the quasi-classical expansion of the linear combination of the transfer matrix of the XXZ Heisenberg spin chain and the central element, the so-called Sklyanin determinant. The corresponding Gaudin Hamiltonians with boundary terms are obtained as the residues of the generating function. By defining the appropriate Bethe vectors which yield strikingly simple off-shell action of the generating function, we fully implement the algebraic Bethe ansatz, obtaining the spectrum of the generating function and the corresponding Bethe equations.
\end{abstract}

\clearpage
\newpage

\section{Introduction \label{sec: intro}}
The so-called rational $s\ell(2)$ Gaudin model was introduced by Gaudin himself as a model of "long-range" interacting spins in a chain \cite{Gaudin76}. This model was extended to any simple Lie algebra, with arbitrary irreducible representation at each site of the chain \cite{Gaudin83,Gaudin-English}. Sklyanin studied the rational model in the framework of the quantum inverse scattering method \cite{TakhtajanFaddeev79,KulishSklyanin82} using the $s\ell(2)$ invariant classical r-matrix \cite{Sklyanin89}. A generalization of these results to all cases when skew-symmetric r-matrix satisfies the classical Yang-Baxter equation \cite{BelavinDrinfeld} was relatively straightforward \cite{SklyaninTakebe,Semenov97}. Therefore, considerable attention has been devoted to Gaudin models corresponding to the the classical r-matrices of simple Lie algebras \cite{Jurco89, Jurco90, BabujianFlume, FeiginFrenkelReshetikhin, ReshetikhinVarchenko, WagnerMacfarlane00} and Lie superalgebras \cite{BrzezinskiMacfarlane94, KulishManojlovic01, KulishManojlovic03, LimaUtiel01, KurakLima04}.

Hikami, Kulish and Wadati showed that the quasi-classical expansion of the  transfer matrix of the periodic chain, calculated at the special values of the spectral parameter, yields the Gaudin Hamiltonians \cite{HikamiKulishWadati92,HikamiKulishWadati92a}. Hikami showed how the quasi-classical expansion of the transfer matrix, calculated at the special values of the spectral parameter, yields the Gaudin Hamiltonians in the case of non-periodic boundary conditions \cite{Hikami95}. Then the ABA was applied to open Gaudin model in the context of the the Vertex-IRF correspondence \cite{YangZhangSasakic04, YangZhangSasakic05, YangZhang12}. Also,  results were obtained for the open Gaudin models based on Lie superalgebras \cite{Lima09}. An approach to study the open Gaudin models based on the classical reflection equation \cite{Sklyanin86,Sklyanin87,Sklyanin88} and the non-unitary r-matrices was developed recently, see \cite{Skrypnyk09, Skrypnyk13} and the references therein. For a review of the open Gaudin model see \cite{CAMN}. Progress in applying Bethe ansatz to the Heisenberg spin chain with non-periodic boundary conditions compatible with the integrability of the quantum systems \cite{China03, Nepomechie04, French06, Martins05, LucEricRafael07,YangWang13, YangWang13a, Eric2013RMP, Eric13, Belliard13, Lima13,Belliard15,BelliardPimenta15,AvanEtal15,Nepomechie15,China15,CAMS,
NMIS} had recent impact on the study of the corresponding Gaudin model \cite{Inna14,CAMRS}. The so-called $T-Q$ approach to implementation of  Bethe ansatz \cite{YangWang13, YangWang13a} was used to obtain the eigenvalues of the associated Gaudin Hamiltonians and the corresponding Bethe ansatz equations \cite{HaoCaoYang14}. In \cite{NMIS} the off-shell action of the generating function of the trigonometric Gaudin Hamiltonians with boundary terms on the Bethe vectors was obtained through the so-called quasi-classical limit.

This work is centered on two objectives, the derivation of the generating function of the Gaudin Hamiltonians and  the implementation of the algebraic Bethe ansatz for the trigonometric Gaudin model, with triangular reflection matrix. In the derivation of the generating function of the Gaudin Hamiltonians with boundary terms we follow the same approach used previously in the rational case \cite{CAMRS}, which in turn was based on Sklyanin's method in the periodic case \cite{Sklyanin89,MNS}. Our derivation is based on the quasi-classical expansion of the linear combination of the transfer matrix of the XXZ Heisenberg spin chain and the central element, the so-called Sklyanin determinant. The essential step in this derivation is the expansion of the monodromy matrix in powers of the quasi-classical parameter. The Gaudin Hamiltonians with the boundary terms are obtained from the residues of the generating function at poles.

The implementation of the algebraic Bethe ansatz in this case is based on the linear bracket defined using the non-unitary classical r-matrix, which satisfies the generalized classical Yang-Baxter equation, and the corresponding modified Lax matrix. This linear bracket is anti-symmetric and it obeys the Jacobi identity. We will show how a suitable set of generators simplifies the commutation relations and facilitates the algebraic Bethe ansatz. Probably the simplest way to define the Bethe vectors is through the family of the creation operators $\mathcal{C}_K(\mu)$. These Bethe vectors $\varphi _M (\mu _1,  \mu_2 , \ldots,\mu_M)$ are symmetric functions of their arguments and they yield strikingly simple off-shell action of the generating function. Actually, it is as simple as it can be since it  practically coincide with the corresponding formula in the case when the boundary matrix is diagonal \cite{Hikami95}. The off-shell action of the generating function of the Gaudin Hamiltonians with the boundary terms yields the spectrum of the system and the corresponding Bethe equations. As usual, when the Bethe equations are imposed on the parameters of the Bethe vectors, the unwanted terms in the action of the generating function are annihilated.

This paper is organized as follows. In Section \ref{sec: XXZ-chain} we review some properties of the Lax operator and other fundamental tools in the study of the inhomogeneous XXZ Heisenberg spin chain,  including the general solutions of the reflection equation and the dual reflection equation. As one of the main results of the paper,  the generating function of the Gaudin Hamiltonians with boundary terms is derived in Section \ref{sec: trig-Gm}, using the quasi-classical expansion of the linear combination of the transfer matrix of the inhomogeneous XXZ spin chain and the so-called Sklyanin determinant. The relevant algebraic structure as well as the implementation of the algebraic Bethe ansatz are presented in Section \ref{sec: lbr-aba}, including the definition of the Bethe vectors and the formulae of the off-shell action of the generating function of the Gaudin Hamiltonians. Our conclusions are presented in the Section \ref{sec: Conclu}. Finally, in Appendix \ref{app: basic-def} are given some basic definitions for the convenience of the reader and some explicit formulas regarding the Bethe vector $\varphi _3 (\mu _1, \mu _2, \mu _3)$ are given in the Appendix \ref{app: phi3}.

\section{Inhomogeneous XXZ  Heisenberg spin chain \label{sec: XXZ-chain}}
We briefly describe the inhomogeneous XXZ Heisenberg spin chain with $N$ sites, characterised by the local space $V_ m = \mathbb{C}^{2s+1}$ and inhomogeneous parameter $\alpha _m$. The Hilbert space of the system is
\begin{equation}
\label{eq: H-space}
\mathcal{H} = \underset {m=1}{\overset {N}{\otimes}}  V_m = (\mathbb{C}^{2s+1} ) ^{\otimes N}.
\end{equation}
Here we follow \cite{NMIS} closely and set the Lax operator to be
\begin{equation}
\label{eq: L-XXZ}
\mathbb{L}_{0m}(\lambda)  
%= \mathbbm{1} + \frac{\eta}{\lambda} \left( \vec{\sigma}_{0} \cdot \vec{S}_{m} \right) 
= \frac{1}{\sinh (\lambda)} \left(\begin{array}{cc}  
\sinh \left( \lambda \mathbbm{1} _m + \eta S_m^{3} \right) & \sinh (\eta)  S_m^{-} \\[1ex] 
\sinh (\eta) S_m^{+} & \sinh \left( \lambda \mathbbm{1} _m - \eta S_m^{3} \right) 
\end{array}\right).
\end{equation}
Evidently this Lax operator is a two-by-two matrix in the auxiliary space $V_0 = \mathbb{C}^2$ and it obeys
\begin{equation}
\label{eq: unitarity-Lax}
\mathbb{L}_{0m}(\lambda) \mathbb{L}_{0m}(\eta-\lambda) = \frac{\sinh \left( s_m \eta +\lambda \right) \sinh\left((s_m + 1) \eta - \lambda \right) }{\sinh (\lambda) \sinh (\eta - \lambda)}
\mathbbm{1} _0 \,,
\end{equation}
where $s_m$ is the value of spin in the space $V_m$ \cite{NMIS}. The  R-matrix of the XXZ  Heisenberg spin chain is given by
\begin{equation}
\label{eq: XXZ-Rmatrix}
R (\lambda, \eta) = \left(\begin{array}{cccc}
\sinh(\lambda + \eta) & 0 & 0 & 0 \\[1ex]
0 & \sinh(\lambda) & \sinh(\eta) & 0 \\[1ex]
0 & \sinh(\eta) & \sinh(\lambda) & 0 \\[1ex]
0 & 0 & 0 & \sinh(\lambda + \eta) \end{array}\right) ,
\end{equation}
where $\lambda$ is a spectral parameter, $\eta$ is a quasi-classical parameter. Due to the commutation relations \eqref{eq: cr-spin-s} it is straightforward to check the RLL-relations
\begin{equation}
\label{eq: RLL}
R_{00'} ( \lambda - \mu) \mathbb{L}_{0m}( \lambda ) \mathbb{L}_{0'm}( \mu ) =  \mathbb{L}_{0'm}( \mu ) \mathbb{L}_{0m}( \lambda )R_{00'} ( \lambda - \mu).
\end{equation}
%From \eqref{eq: unitarity-Lax} it follows
%\begin{equation}
%\label{eq: LRL}
%\mathbb{L}_{0'm}( \eta - \mu ) R_{00'} ( \lambda - \mu) \mathbb{L}_{0m}( \lambda ) = \mathbb{L}_{0m}( \lambda )R_{00'} ( \lambda - \mu) \mathbb{L}_{0'm}( \eta - \mu ).
%\end{equation}
%Analogously,  
%\begin{equation}
%\label{eq: LLR}
%\mathbb{L}_{0m}( \eta - \lambda ) \mathbb{L}_{0'm}( \eta - \mu ) R_{00'} ( \lambda - \mu)  = R_{00'} ( \lambda - \mu) \mathbb{L}_{0'm}( \eta - \mu ) \mathbb{L}_{0m}( \eta - \lambda ).
%\end{equation}

In this brief description of the XXZ Heisenberg spin chain with non-periodic boundary condition, we will follow Sklyanin's approach \cite{Sklyanin88}.
Boundary conditions on the left and right sites of the chain are encoded in the left and right reflection matrices $K^-$ and $K^+$. The compatibility condition between the bulk and the boundary of the system takes the form of the so-called reflection equation. It is written in the following form for the left reflection matrix $K^-(\lambda) \in \mathrm{End} (\mathbb{C}^2)$
\begin{equation}
\label{eq: RE}
R_{12}(\lambda - \mu) K^-_1(\lambda) R_{21}(\lambda + \mu) K^-_2(\mu)=
K^-_2(\mu) R_{12}(\lambda + \mu) K^-_1(\lambda) R_{21}(\lambda - \mu) .
\end{equation}
The general, spectral parameter dependent, solutions of the reflection equation \eqref{eq: RE} can be written as follows \cite{VegaGonzalez, Zamolodchikov94,Zamolodchikov94b}
\begin{equation}
\label{eq: K-min}
K ^{-}(\lambda) =  \left(\begin{array}{cc}
\kappa ^{-} \sinh (\xi ^{-} + \lambda) & \psi ^{-} \sinh (2 \lambda) \\ 
\phi ^{-} \sinh (2 \lambda) & \kappa ^{-}\sinh (\xi ^{-} - \lambda) \end{array}\right) .
\end{equation}
%satisfies the reflection equation
%\begin{equation}
%\label{eq: RE}
%R_{12}(\lambda - \mu) K^-_1(\lambda) R_{21}(\lambda + \mu) K^-_2(\mu)=
%K^-_2(\mu) R_{12}(\lambda + \mu) K^-_1(\lambda) R_{21}(\lambda - \mu) .
%\end{equation}

The main tool in Sklyanin's framework is the corresponding monodromy matrix  
\begin{equation}
\label{eq: calT}
\mathcal{T}_0(\lambda)= T_0(\lambda) K _0^{-}(\lambda) \widetilde T_0(\lambda),
\end{equation}
it consists of the reflection matrix $K_0 ^{-}(\lambda)$ and the two monodromy matrices $T_0 (\lambda)$ and $\widetilde{T}_0(\lambda )$
\begin{align}
\label{eq: mono-T}
T_0(\lambda ) &= \
\mathbb{L}_{0N} ( \lambda - \alpha _N) \cdots \mathbb{L}_{01} ( \lambda - \alpha _1) , \\
\label{eq: mono-tilde-T}    
\widetilde{T}_0(\lambda ) &= 
\mathbb{L}_{01}(\lambda + \alpha _1 + \eta) \cdots \mathbb{L}_{0N} (\lambda + \alpha _N + \eta) .
\end{align}
From the definitions \eqref{eq: mono-T} and \eqref{eq: mono-tilde-T} and the relations \eqref{eq: unitarity-Lax} and \eqref{eq: RLL} follow the RTT-relations 
\begin{align}
\label{eq: RTT}
R_{00'} ( \lambda - \mu) {T}_{0} (\lambda ) {T}_ {0'}(\mu ) &=  {T} _ {0'}(\mu ){T}_ {0}(\lambda ) R_{00'} ( \lambda - \mu) , \\
\label{eq: tTRT}
\widetilde{T} _ {0'}(\mu ) R_{00'} ( \lambda + \mu) T _{0} (\lambda )  &= T _ {0}(\lambda ) R_{00'} ( \lambda + \mu) \widetilde{T} _ {0'}(\mu ) , \\
\label{eq: tTtTR}
\widetilde{T} _{0} (\lambda ) \widetilde{T} _ {0'}(\mu ) R_{00'} (\mu - \lambda) &= R_{00'} (\mu - \lambda) \widetilde{T} _ {0'}(\mu ) \widetilde{T} _ {0}(\lambda ) .
\end{align}
Using these RTT-relations \eqref{eq: RTT}, \eqref{eq: tTRT}, \eqref{eq: tTtTR} and the reflection equation \eqref{eq: RE} it is straightforward to show that the exchange relations of the monodromy matrix $\mathcal{T}(\lambda)$  in $V_0\otimes V_{0'}$ are \cite{Sklyanin88}
\begin{equation}
\label{eq: exchangeRE}
R _{00'}(\lambda - \mu) \mathcal{T}_{0} (\lambda) R _{0'0} (\lambda + \mu) \mathcal{T} _{0'} (\mu) = 
\mathcal{T}_{0'}(\mu) R _{00^{\prime}} (\lambda + \mu) \mathcal{T}_{0} (\lambda) 
R _{0'0} (\lambda - \mu) .
\end{equation}
These exchange relations \eqref{eq: exchangeRE} admit a central element,  the so-called Sklyanin determinant,
\begin{equation}
\label{eq: Delta-T-cal}
\Delta \left[\mathcal{T}(\lambda)\right] = \mathrm{tr}_{00'} P^{-}_{00'} \mathcal{T}_{0}(\lambda-\eta/2) R_{00'} (2\lambda) \mathcal{T}_{0'}(\lambda+\eta/2). 
\end{equation}

The open chain transfer matrix is given by the trace of the monodromy $\mathcal{T}(\lambda)$ over the auxiliary space $V_0$ with an extra reflection matrix $K^+(\lambda)$ \cite{Sklyanin88},
\begin{equation}
\label{eq: open-t}
t (\lambda) = \mathrm{tr}_0 \left( K^+(\lambda) \mathcal{T}(\lambda) \right).
\end{equation}
The reflection matrix $K^+(\lambda)$ %\eqref{K-plus} 
is the corresponding solution of the dual reflection equation \cite{NMIS}
\begin{equation}
\label{eq: Kpl}
K^+(\lambda)= K^{-}(- \lambda -\eta) =
\left(\begin{array}{cc}
\kappa ^{+} \sinh ( \xi ^{+} - \lambda - \eta) & - \psi ^{+} \sinh \left( 2 (\lambda + \eta) \right) \\ 
- \phi ^{+} \sinh \left( 2 (\lambda + \eta) \right) & \kappa ^{+}\sinh (\xi ^{+} + \lambda + \eta)
\end{array}\right) .
\end{equation}
Due to the fact that the reflection matrices $K ^{\mp}(\lambda)$ are defined up to multiplicative constants the values of parameters $\kappa ^{\mp}$ are not essential, as long as they are different from zero. Therefore they could be set to be one without any loss of generality. In particular, this will be evident throughout the Section \ref{sec: lbr-aba}. However, for completeness, we will keep them in our presentation. 

\section{Trigonometric Gaudin model with boundary \label{sec: trig-Gm}}
The study of the open Gaudin model requires the following condition to be imposed on the reflection matrices \cite{CAMN, CAMRS}
\begin{equation}
\label{eq: normalizationKpl}
\lim_{\eta \to 0}\Big(  K^+(\lambda) K^{-} (\lambda)\Big)  = \left( \kappa ^2 \sinh (\xi - \lambda) \sinh (\xi + \lambda) - \phi \psi \sinh ^2 (\lambda) \right) \mathbbm{1}.
\end{equation}
In particular, this implies that the parameters of the reflection matrices on the left and on the right end of the chain are the same. In general this is not the case in  the study of the open spin chain. However, this condition is essential for the Gaudin model. Then we will write
\begin{equation}
\label{eq: K-min-Gm}
K^-(\lambda)\equiv K(\lambda) = \left(\begin{array}{cc}
\kappa  \sinh (\xi  + \lambda) & \psi  \sinh (2 \lambda) \\ 
\phi  \sinh (2 \lambda) & \kappa \sinh (\xi  - \lambda) \end{array}\right) 
\end{equation}
so that 
\begin{equation} 
\label{eq: K-plus-Gm}
K^+(\lambda)= K(-\lambda-\eta)= \left(\begin{array}{cc}
\kappa \sinh ( \xi - \lambda - \eta) & - \psi  \sinh \left( 2 (\lambda + \eta) \right) \\ 
- \phi \sinh \left( 2 (\lambda + \eta) \right) & \kappa \sinh (\xi  + \lambda + \eta)
\end{array}\right) .
\end{equation}
Moreover, it is straightforward to check the following useful identities
\begin{align}
\label{eq: KK-l}
K (- \lambda)  K (\lambda) &= \mathrm{det} \left(K(\lambda) \right)  \mathbbm{1} , \\
\label{eq: Kmin-l}    
K (- \lambda) &= \mathrm{tr} \ K(\lambda) - K(\lambda) .       
\end{align}

%With the aim of obtaining the expansion in $\eta$ the Lax operator \eqref{eq: L-XXZ} is given as follows
%\begin{equation}
%\label{L-XXZ-eta}
%\mathbb{L}_{0m}(\lambda - \alpha _m )   
%=  \mathbbm{1} _0 \otimes \cosh (\eta S_m^3) + \coth (\lambda - \alpha _m ) \ \sigma _0^3 \otimes \sinh (\eta S_m^3) + \frac{\sinh (\eta)}{2 \ \sinh (\lambda - \alpha _m )} \left( \sigma _0^+ \otimes S_m^- + \sigma _0^- \otimes S_m^+ \right) .
%\end{equation}
Analogously to the rational case \cite{CAMRS}, our first step in the derivation of the Gaudin model is to obtain first few terms in the  power series in $\eta$ of the Lax operator \eqref{eq: L-XXZ} 
\begin{equation}
\label{L-XXZ-eta-series}
\begin{split}
\mathbb{L}_{0m}(\lambda - \alpha _m )   
&=  \mathbbm{1} _0 \otimes \mathbbm{1} _m + \eta \left( \sigma _0^3 \otimes \  \coth (\lambda - \alpha _m ) \ S_m^3 + \frac{1}{2 \ \sinh (\lambda - \alpha _m )}  \left( \sigma _0^+ \otimes S_m^- + \sigma _0^- \otimes S_m^+ \right) \right) \\
&+ \frac{\eta^2}{2} \mathbbm{1} _0 \otimes \left( S_m^3 \right) ^2  + \mathcal{O}(\eta ^3) .
\end{split}
\end{equation}
It is important to notice that the spin operators $S_m^{\alpha}$, with $\alpha = +, - , 3$, on the right hand side of  \eqref{L-XXZ-eta-series} satisfy the usual commutation relations
\begin{equation}
\label{crspin1}
[S_m^3, S_n^{\pm}] = \pm S_m^{\pm} \, \delta_{mn}, \quad [S_m^+,S_n^-] = 2 S_m^3 \, \delta_{mn} . 
\end{equation}
As expected, the term linear in $\eta$ in the expansion above defines the Gaudin Lax matrix \cite{NMIS, MNS}
\begin{equation}
\label{L-Gaudin}
L_0 (\lambda) = \sum _{m=1}^N  \frac{\sigma_{0}^3 \otimes \cosh (\lambda - \alpha _m) S_{m}^3 + \frac{1}{2} \left( \sigma_{0}^+ \otimes S_{m}^- + \sigma_{0}^- \otimes S_{m}^+ \right)}{\sinh (\lambda - \alpha _m)} .
\end{equation}
When the quasi-classical property of the R-matrix \eqref{eq: XXZ-Rmatrix} %\cite{KulishSklyanin82}
\begin{equation}
\label{quasi-classical}
\frac{1}{\sinh (\lambda)} R (\lambda) = \mathbbm{1} - \eta r(\lambda) + \mathcal{O}(\eta ^2), \end{equation}
where
\begin{equation}
\label{eq: classical-r}
r(\lambda) = \frac{-1}{2\sinh (\lambda)} \left( \cosh (\lambda) (\mathbbm{1} \otimes \mathbbm{1} + \sigma ^3 \otimes \sigma ^3) + \frac{1}{2} \left( \sigma ^+ \otimes \sigma ^- + \sigma ^- \otimes \sigma^+ \right) \right) ,
\end{equation}
is taken into account, then substitution of the expansion \eqref{L-XXZ-eta-series} into the RLL-relations \eqref{eq: RLL} yields the so-called Sklyanin linear bracket %\cite{Sklyanin89}
\begin{equation}
\label{rLL}
\left[ L _1(\lambda) , L _2 (\mu) \right] = \left[ r_{12}(\lambda - \mu) , L _1(\lambda) + L _2 (\mu) \right].
\end{equation}
The classical r-matrix \eqref{eq: classical-r} has the unitarity property 
\begin{equation}
\label{eq: r-unitarity}
r_{21}(-\lambda) = - r_{12}(\lambda) ,
\end{equation}
and satisfies the classical Yang-Baxter equation %\cite{BelavinDrinfeld}
\begin{equation}
\label{eq: classicalYBE}
[r_{13} (\lambda), r_{23}(\mu) ] + [r_{12}(\lambda - \mu), r_{13} (\lambda) +  r_{23}(\mu)] =0.
\end{equation}
Thus the Sklyanin linear bracket \eqref{rLL} is anti-symmetric and it obeys the Jacobi identity. It follows that the entries of the Lax matrix \eqref{L-Gaudin} generate a Lie algebra, the so-called Gaudin algebra, in particular, relevant for the trigonometric Gaudin model with periodic boundary conditions \cite{KulishManojlovic03}.

Our next step is to consider the expansion of the monodromy matrix \eqref{eq: mono-T} with respect to the quasi-classical parameter $\eta$
\begin{align}
\label{eq: expan-T}
T(\lambda ) &= \mathbbm{1} + \eta L_0 (\lambda) + \frac{\eta ^2}{2} \mathbbm{1}_0 \otimes \sum _{m=1}^N  \left( S_{m}^3 \right) ^2
\notag \\
&+ \frac{\eta ^2}{2} \underset {n \neq m} {\sum _{n,m=1}^N} \frac{\mathbbm{1}_0 \otimes \left( \cosh (\lambda - \alpha _m) \cosh (\lambda - \alpha _n) \, S_{m}^3 S_{n}^3 + \frac{1}{2} \left( S_{m}^+ S_{n}^- + S_{m}^- S_{n}^+  \right) \right)}{\sinh (\lambda - \alpha _m) \ \sinh (\lambda - \alpha _n)} 
\notag \\
&+  \frac{\eta ^2}{2} \sum _{m=1}^N  \sum _ {n < m}^N \frac{ \sigma_{0}^3 \otimes \left( S_{m}^- S_{n}^+ - S_{m}^+ S_{n}^-  \right) + \sigma_{0}^+ \otimes \left( \cosh (\lambda - \alpha _m) S_{m}^3 S_{n}^- -  \cosh (\lambda - \alpha _n) S_{m}^- S_{n}^3 \right) } {2 \sinh(\lambda - \alpha _m) \ \sinh (\lambda - \alpha _n)} 
\notag \\
&+ \frac{\eta ^2}{2} \sum _{m=1}^N  \sum _ {n < m}^N \frac{ \sigma_{0}^- \otimes  \left( \cosh (\lambda - \alpha _n) S_{m}^+ S_{n}^3 -  \cosh (\lambda - \alpha _m) S_{m}^3 S_{n}^- \right) 
}{2 \sinh(\lambda - \alpha _m) \ \sinh (\lambda - \alpha _n)} 
\notag \\
&+  \frac{\eta ^2}{2} \sum _{m=1}^N \sum _{n > m}^N \frac{ \sigma_{0}^3 \otimes \left( S_{n}^- S_{m}^+ - S_{n}^+ S_{m}^-  \right) + \sigma_{0}^+ \otimes \left( \cosh (\lambda - \alpha _n) S_{n}^3 S_{m}^- -  \cosh (\lambda - \alpha _m) S_{n}^- S_{m}^3 \right) } {2 \sinh(\lambda - \alpha _n) \ \sinh (\lambda - \alpha _m)} 
\notag \\
&+  \frac{\eta ^2}{2} \sum _{m=1}^N \sum _{n > m}^N \frac{ \sigma_{0}^- \otimes  \left( \cosh (\lambda - \alpha _m) S_{n}^+ S_{m}^3 -  \cosh (\lambda - \alpha _n) S_{n}^3 S_{m}^- \right) 
}{2 \sinh(\lambda - \alpha _n) \ \sinh (\lambda - \alpha _m)} 
+ \mathcal{O}(\eta ^3) .
\end{align}

%Taking into account that from \eqref{L-XXZ-eta} it follows that
%\begin{equation}
%\label{L-XXZ-eta-l}
%\begin{split}
%\mathbb{L}_{0m}(\lambda + \alpha _m + \eta )   
%&=  \mathbbm{1} _0 \otimes \cosh (\eta S_m^3) + \coth (\lambda + \alpha _m + \eta ) \ \sigma _0^3 \otimes \sinh (\eta S_m^3) \\ 
%%
%&+ \frac{\sinh (\eta)}{2 \ \sinh (\lambda + \alpha _m + \eta)} \left( \sigma _0^+ \otimes S_m^- + \sigma _0^- \otimes S_m^+ \right) ,
%\end{split}
%\end{equation}
%and therefore
%\begin{equation}
%\label{L-XXZ-l-eta-series}
%\begin{split}
%\mathbb{L}_{0m}(\lambda + \alpha _m + \eta )   
%&=  \mathbbm{1} _0 \otimes \mathbbm{1} _m + \eta \left( \sigma _0^3 \otimes \  \coth (\lambda + \alpha _m ) \ S_m^3 + \frac{1}{2 \ \sinh (\lambda + \alpha _m )}  \left( \sigma _0^+ \otimes S_m^- + \sigma _0^- \otimes S_m^+ \right) \right) \\
%&+ \frac{\eta^2}{2} \left( \mathbbm{1} _0 \otimes ( S_m^3 )^2  - \frac{2 \left( \sigma^3 _0 \otimes S_m^3 
%+ \frac{1}{2} \cosh (\lambda + \alpha _m ) \left(\sigma _0^+ \otimes S_m^- + \sigma _0^- \otimes S_m^+ \right) \right)}{\sinh ^2 (\lambda + \alpha _m) } \right) \\
%%
%&+ \mathcal{O}(\eta ^3) ,
%\end{split}
%\end{equation}
Analogously, it is straightforward to obtain the expansion of the monodromy matrix \eqref{eq: mono-tilde-T} in the powers the quasi-classical parameter $\eta$
\begin{align}
\label{eq: expan-tilde-T}
\widetilde{T}(\lambda ) &= \mathbbm{1} - \eta L_0 ( - \lambda) + \frac{\eta ^2}{2} \sum _{m=1}^N  \left(\mathbbm{1} _0 \otimes ( S_m^3 )^2  - \frac{2 \left( \sigma^3 _0 \otimes S_m^3 
+ \frac{1}{2} \cosh (\lambda + \alpha _m ) \left(\sigma _0^+ \otimes S_m^- + \sigma _0^- \otimes S_m^+ \right) \right)}{\sinh ^2 (\lambda + \alpha _m) } \right) 
\notag \\
&+ \frac{\eta ^2}{2} \underset {n \neq m} {\sum _{n,m=1}^N} \frac{\mathbbm{1}_0 \otimes \left( \cosh (\lambda + \alpha _m) \cosh (\lambda + \alpha _n) \, S_{m}^3 S_{n}^3 + \frac{1}{2} \left( S_{m}^+ S_{n}^- + S_{m}^- S_{n}^+  \right) \right)}{\sinh (\lambda + \alpha _m) \ \sinh (\lambda + \alpha _n)} 
\notag \\
&+  \frac{\eta ^2}{2} \sum _{m=1}^N  \sum _ {n < m}^N \frac{ \sigma_{0}^3 \otimes \left( S_{m}^- S_{n}^+ - S_{m}^+ S_{n}^-  \right) + \sigma_{0}^+ \otimes \left( \cosh (\lambda + \alpha _m) S_{m}^3 S_{n}^- -  \cosh (\lambda + \alpha _n) S_{m}^- S_{n}^3 \right) } {2 \sinh(\lambda + \alpha _m) \ \sinh (\lambda + \alpha _n)} 
\notag \\
&+ \frac{\eta ^2}{2} \sum _{m=1}^N  \sum _ {n < m}^N \frac{ \sigma_{0}^- \otimes  \left( \cosh (\lambda - \alpha _n) S_{m}^+ S_{n}^3 -  \cosh (\lambda + \alpha _m) S_{m}^3 S_{n}^- \right) 
}{2 \sinh(\lambda + \alpha _m) \, \sinh (\lambda + \alpha _n)} 
\notag \\
&+  \frac{\eta ^2}{2} \sum _{m=1}^N \sum _{n > m}^N \frac{ \sigma_{0}^3 \otimes \left( S_{n}^- S_{m}^+ - S_{n}^+ S_{m}^-  \right) + \sigma_{0}^+ \otimes \left( \cosh (\lambda + \alpha _n) S_{n}^3 S_{m}^- -  \cosh (\lambda + \alpha _m) S_{n}^- S_{m}^3 \right) } {2 \sinh(\lambda + \alpha _n) \, \sinh (\lambda - \alpha _m)} 
\notag \\
&+  \frac{\eta ^2}{2} \sum _{m=1}^N \sum _{n > m}^N \frac{ \sigma_{0}^- \otimes  \left( \cosh (\lambda + \alpha _m) S_{n}^+ S_{m}^3 -  \cosh (\lambda + \alpha _n) S_{n}^3 S_{m}^- \right) 
}{2 \sinh(\lambda + \alpha _n) \, \sinh (\lambda + \alpha _m)} + \mathcal{O}(\eta ^3) .
\end{align}

%Finally, form \eqref{eq: K-plus-Gm}, it follows that
%\begin{equation}
%\label{eq: expan-K-plus-Gm}
%K (-\lambda -\eta) = K (-\lambda ) + \eta \ K^{\prime} (-\lambda ) + \frac{\eta ^2}{2} \ K^{\prime \prime} (-\lambda ) + \mathcal{O}(\eta ^3) ,
%\end{equation}
%where 
%\begin{equation}
%     \frac{d K (-\lambda - \eta )}{d\eta} \Big{|}_{\eta=0}  =  (-1) K^{\prime} (-\lambda ) , \quad \text{and} \quad
%     \frac{d^2K (-\lambda - \eta )}{d\eta ^2} \Big{|}_{\eta=0}  = K^{\prime \prime} (-\lambda ) .  
%\end{equation}

Using the formulae above \eqref{eq: expan-T} and \eqref{eq: expan-tilde-T} as well as the first three terms in the power series of the K-matrix \eqref{eq: K-plus-Gm} we can deduce the expansion of $t (\lambda)$ \eqref{eq: open-t} in powers of $\eta$.  Similarly, the expansion of $\Delta \left[\mathcal{T}(\lambda) \right]$ \eqref{eq: Delta-T-cal} in powers of $\eta$ is obtained. However these formulas are long and cumbersome, therefore we will not present them here. Instead we will give the expansion of the difference between the transfer matrix of the chain and the so-called Sklyanin determinant. In order to simplify these formulae we introduce the following notation
\begin{equation}
\label{eq: cal-L}
\mathcal{L} _0 (\lambda) = L _0 (\lambda) - K_0 (\lambda)  L _0 (- \lambda) K_0^{-1}(\lambda), \\
\end{equation}
where the Gaudin Lax matrix $L_0 (\lambda)$ and the reflection matrix $K_0 (\lambda)$ are given in \eqref{L-Gaudin} and \eqref{eq: K-min-Gm}, respectively. Therefore, it can be shown that, using the notation above, the expansion in powers of $\eta$ of the difference between $t (\lambda)$ and $\Delta \left[\mathcal{T}(\lambda) \right]$ is given by
\begin{align}
\label{expan-open-t-Delta-cal-T}
& t (\lambda) - \frac{\Delta \left[\mathcal{T}(\lambda) \right] }{\sinh (2\lambda)} = \frac{1}{2}  \mathrm{tr}_0   K _0 (\lambda)  K _0 (-\lambda) + \eta \left( \mathrm{tr}_0 K _0^{\prime} (- \lambda) K _0 (\lambda) + \mathrm{tr}_{00'} P^{-}_{00'} K _0 (\lambda) r_{00'} (2\lambda) K _{0'} (\lambda) \right)
\notag \\[1ex]
&+ \eta ^2 \left( \mathrm{tr}_0 K _0^{\prime} (- \lambda) \mathcal{L} _0 (\lambda) K _0 (\lambda) +  
\mathrm{tr}_{00'} P^{-}_{00'} \left( \mathcal{L} _0 (\lambda) K _0 (\lambda) r_{00'} (2\lambda) K _{0'} (\lambda) + K _0 (\lambda) r_{00'} (2\lambda) \mathcal{L} _{0'} (\lambda) K _{0'} (\lambda) \right) \right)
\notag \\[1ex]
&- \eta ^2 \ \mathrm{tr}_{00'} P^{-}_{00'} \mathcal{L} _0 (\lambda) K _0 (\lambda)  \mathcal{L} _{0'} (\lambda) K _{0'} (\lambda) + \frac{\eta ^2}{2} \left( \mathrm{tr}_0 K _0^{\prime\prime} (- \lambda) K _0 (\lambda) - \frac{1}{4} \ \mathrm{tr}_0 K _0^{\prime\prime} (\lambda) K _0 (-\lambda) \right.
\notag \\[1ex]
&\left. + \frac{1}{2} \  \mathrm{tr}_{00'} P^{-}_{00'} K _0^{\prime} (\lambda) K _{0'}^{\prime} (\lambda)  - \frac{1}{\sinh (2\lambda)} \ \mathrm{tr}_{00'} P^{-}_{00'} K _0 (\lambda) \partial _{\eta}^2 R_{00'} (2\lambda) \big{|}_{\eta = 0} K _{0'} (\lambda) \right) + \mathcal{O}(\eta ^3) .
\end{align}
Actually, a straightforward calculation shows that the terms in the second line of the expression above vanish
\begin{equation}
\label{identity-1}
\mathrm{tr}_0 K _0^{\prime} (- \lambda) \mathcal{L} _0 (\lambda) K _0 (\lambda) +  
\mathrm{tr}_{00'} P^{-}_{00'} \left( \mathcal{L} _0 (\lambda) K _0 (\lambda) r_{00'} (2\lambda) K _{0'} (\lambda) + K _0 (\lambda) r_{00'} (2\lambda) \mathcal{L} _{0'} (\lambda) K _{0'} (\lambda) \right) = 0 .
\end{equation}
Also, it is important to notice that using the following identity
\begin{equation}
\label{mat-id}
\mathcal{L}_0 (\lambda) K_0 (\lambda) - \mathrm{tr}_{0'} \left(  \mathcal{L}_{0'} (\lambda) K_{0'} (\lambda) \right) \mathbbm{1}_0 = K_0 (- \lambda) \mathcal{L}_0 (\lambda) ,
\end{equation}
the first term in the third line of \eqref{expan-open-t-Delta-cal-T} can be simplified 
\begin{equation}
\label{ 3-rd-term-sim}
\mathrm{tr}_0 \, K_0 (- \lambda) \mathcal{L}_0 (\lambda) \mathcal{L}_0 (\lambda) K_0 (\lambda) =  \det   K_0 (\lambda) \ \mathrm{tr}_0 \, \mathcal{L}_0 ^2(\lambda) .
\end{equation}
Finally, the expansion \eqref{expan-open-t-Delta-cal-T} reads
\begin{align}
\label{final-expan-open-t-Delta-cal-T}
t (\lambda) - \frac{\Delta \left[\mathcal{T}(\lambda) \right] }{\sinh (2\lambda)}  &= \det K_0 (\lambda) + \eta \left( \mathrm{tr}_0 K _0^{\prime} (- \lambda) K _0 (\lambda) + \mathrm{tr}_{00'} P^{-}_{00'} K _0 (\lambda) r_{00'} (2\lambda) K _{0'} (\lambda) \right)
\notag \\[1ex]
&+ \frac{\eta ^2}{2} \ \det K_0 (\lambda) \ \mathrm{tr}_0 \, \mathcal{L}_0 ^2(\lambda)  + \frac{\eta ^2}{2} \left( \mathrm{tr}_0 K _0^{\prime\prime} (- \lambda) K _0 (\lambda) - \frac{1}{4} \ \mathrm{tr}_0 K _0^{\prime\prime} (\lambda) K _0 (-\lambda) \right.
\notag \\[1ex]
&\left. + \frac{1}{2} \  \mathrm{tr}_{00'} P^{-}_{00'} K _0^{\prime} (\lambda) K _{0'}^{\prime} (\lambda)  - \frac{1}{\sinh (2\lambda)} \ \mathrm{tr}_{00'} P^{-}_{00'} K _0 (\lambda) \partial _{\eta}^2 R_{00'} (2\lambda) \big{|}_{\eta = 0} K _{0'} (\lambda) \right) + \mathcal{O}(\eta ^3) .
\end{align}
As expected \cite{CAMRS}, this shows that
\begin{equation}
\label{eq: b-tau} 
\tau (\lambda) =  \mathrm{tr}_0 \, \mathcal{L}_0 ^2(\lambda) 
\end{equation}
commutes for different values of the spectral parameter,
\begin{equation}
\label{eq: b-tau-tau}
[ \tau (\lambda) , \tau (\mu) ] = 0.
\end{equation}
Therefore $\tau (\lambda)$ \eqref{eq: b-tau} can be considered to be the generating function of Gaudin Hamiltonians with boundary terms. In the next section we will obtain these Gaudin Hamiltonians explicitly as well as the spectrum and the corresponding Bethe vectors of the generating function.

\section{Linear bracket and algebraic Bethe ansatz \label{sec: lbr-aba}}
With the aim of implementing the algebraic Bethe ansatz to the trigonometric Gaudin model with triangular K-matrix we seek the linear bracket relations for the Lax operator \eqref{eq: cal-L}. As in the rational case \cite{CAMRS}, the classical r-matrix \eqref{eq: classical-r} satisfies the classical Yang-Baxter equation \eqref{eq: classicalYBE} and has the unitarity property \eqref{eq: r-unitarity}. Moreover, the classical r-matrix and the reflection matrix \eqref{eq: K-min-Gm} satisfy the classical reflection equation
\begin{equation}
\label{eq: c-RE}
\begin{split}
r_{12}(\lambda - \mu) K _1(\lambda) K_2(\mu) + K _1(\lambda) r_{21}(\lambda + \mu) K_2(\mu) = \\
= K_2(\mu) r_{12}(\lambda + \mu) K _1(\lambda) +  K_2(\mu) K _1(\lambda) r_{21}(\lambda - \mu) .
\end{split}
\end{equation}
With the aim of obtaining the suitable linear bracket relations of the Lax operator \eqref{eq: cal-L} we define the non-unitary r-matrix 
\begin{equation}
\label{eq: rK}
r_{00'}^{K} (\lambda , \mu) = r_{00'} (\lambda - \mu) - K_{0'} (\mu) r_{00'} (\lambda + \mu) K_{0'}^{-1} (\mu).
\end{equation}
It is straightforward to check that this r-matrix satisfies the classical Yang-Baxter equation
\begin{equation}
\label{eq: g-c-YBE}
[ r^K_{32}(\lambda_3,\lambda_2) , r^K_{13} (\lambda_1,\lambda_3) ] + [r^K_{12}(\lambda_1, \lambda_2), r^K_{13} (\lambda_1,\lambda_3) + r^K_{23}(\lambda_2,\lambda_3) ] 
=0. 
\end{equation}
The linear bracket for the he Lax operator \eqref{eq: cal-L} reads
\begin{equation}
\label{eq: linear-rk-br}
\left[ \mathcal{L} _{0} (\lambda) , \mathcal{L} _{0'}(\mu) \right] = \left[ r_{00'}^{K} (\lambda , \mu) , \mathcal{L} _{0} (\lambda) \right] - \left[ r_{0'0}^{K} (\mu , \lambda) , \mathcal{L} _{0'} (\mu) \right] .
\end{equation}
This linear bracket is obviously anti-symmetric. It obeys the Jacobi identity because the $r$-matrix \eqref{eq: rK} satisfies the classical Yang-Baxter equation.

Implementation of the algebraic Bethe ansatz requires triangularity of the K-matrix \eqref{eq: K-min-Gm}. As opposed to the rational case \cite{CAMRS} were the triangularity of the K-matrix can be guaranteed by the similarity transformation independent of the spectral parameter, in the present case the reflection matrix cannot be brought to the upper triangular form by the $U(1)$ symmetry transformations. On the contrary,  we have to impose an extra condition on the parameters of $K (\lambda)$. By setting $$\phi =0$$ the reflection matrix becomes upper triangular 
\begin{equation}
\label{eq: K-triangular}
K(\lambda) = \left(\begin{array}{cc}
\kappa  \sinh (\xi  + \lambda) & \psi  \sinh (2 \lambda) \\ 
0 & \kappa \sinh (\xi  - \lambda) \end{array}\right) .
\end{equation}
Evidently, the inverse matrix is
\begin{equation}
\label{eq: inv-K-triangular}
K ^{-1}(\lambda) = \frac{1}{\kappa^2  \sinh (\xi  + \lambda)\sinh (\xi  - \lambda)} \left(
\begin{array}{cc}
\kappa  \sinh (\xi  - \lambda) & - \psi  \sinh (2 \lambda) \\ 
0 & \kappa \sinh (\xi  + \lambda) 
\end{array} \right) .
\end{equation}
Direct substitution of the formulae above into \eqref{eq: cal-L},
\begin{equation}
\label{eq: cal-L-local}
\mathcal{L} _{0} (\lambda) = 
\left(\begin{array}{rr} H(\lambda) & F(\lambda) \\ E(\lambda)  & -H(\lambda)\end{array}\right)
= L _0 (\lambda) - K_0 (\lambda)  L _0 (- \lambda) K_0^{-1}(\lambda),
\end{equation}
yields the following local realisation for the entries of the Lax matrix
\begin{align}
\label{gen-E}
E (\lambda) &=  \sum _{m=1}^N \left ( \frac{S ^+_m}{\sinh(\lambda - \alpha _m)} + \frac{\sinh(\xi - \lambda ) \ S ^+_m}{\sinh(\xi + \lambda )\sinh (\lambda + \alpha _m)}
\right) , \\[1ex]
\label{gen-H} 
H (\lambda) &= \sum _{m=1}^N \left ( \coth (\lambda - \alpha _m) \ S ^3_m + \coth (\lambda + \alpha _m) \ S ^3_m + \frac{\psi \sinh (2\lambda) \ S ^+_m }{\kappa \sinh (\xi + \lambda) \sinh (\lambda + \alpha _m)} \right)  , \\[1ex]
\label{gen-F}
F (\lambda) &= \sum _{m=1}^N \left ( \frac{S ^-_m}{\sinh(\lambda - \alpha _m)} + \frac{\sinh(\xi + \lambda ) \ S ^-_m}{\sinh(\xi - \lambda )\sinh (\lambda + \alpha _m)} 
- \frac{2\psi \sinh(2\lambda)}{\kappa \sinh(\xi - \lambda)} \coth(\lambda + \alpha _m)  \ S ^3_m \right. \notag \\
&\left. \qquad \ - \frac{\psi ^2 \sinh ^2 (2\lambda) \ S ^+_m}{\kappa ^2 \sinh(\xi - \lambda) \sinh(\xi + \lambda) \sinh (\lambda + \alpha _m)} \right) .
\end{align}

Similarly, substituting \eqref{eq: K-triangular} and \eqref{eq: inv-K-triangular} together with \eqref{eq: classical-r} into \eqref{eq: rK} we obtain $r_{00'}^{K} (\lambda , \mu)$ explicitly. This non-unitary, classical r-matrix together with the Lax matrix \eqref{eq: cal-L-local} defines the Lie algebra relevant for the open trigonometric Gaudin model. The relation \eqref{eq: linear-rk-br} is a compact matrix form of the following commutation relations for the generators $E (\lambda)$, $H (\lambda)$ and $F (\lambda)$
\begin{align}
\label{eq: com-EE}
\left[ E (\lambda) , E (\mu) \right] &= 0 ,  \\
\label{eq: com-HE}
\left[ H (\lambda) , E (\mu) \right] &= \frac{1}{\sinh(\lambda - \mu)\sinh (\lambda + \mu)} \left( \sinh(2 \lambda) \ E (\mu) - \frac{\sinh (\xi + \lambda)}{\sinh (\xi + \mu)} \sinh (2\mu) \ E (\lambda) \right) ,  \\[1ex]
\label{eq: com-EF}
\left[ E (\lambda) , F (\mu) \right] &=  \frac{2\psi}{\kappa} \coth(\lambda + \mu) \frac{\sinh (2\mu)}{\sinh (\xi - \mu)} \ E (\lambda) + \frac{2}{\sinh(\lambda - \mu)\sinh (\lambda + \mu)} \times \notag \\
&\times \left( \frac{\sinh (\xi + \mu)}{\sinh (\xi + \lambda)} \sinh(2\lambda) \ H (\mu) -  \frac{\sinh( \xi - \lambda)}{\sinh (\xi - \mu)} \sinh (2\mu) \ H (\lambda) \right) ,  \\[1ex]
\label{eq: com-HH}
\left[ H (\lambda) , H (\mu) \right] &= \frac{- \psi}{\kappa \sinh (\lambda + \mu)} \left( \frac{\sinh (2 \lambda)}{\sinh (\xi + \lambda )} \ E (\mu) - \frac{\sinh (2\mu)}{\sinh (\xi + \mu )} \ E (\lambda) \right), \\[1ex]
%\end{align}
%
%\begin{align}
\label{com-HF}
\left[ H (\lambda) , F (\mu) \right] &= -  \frac{1}{\sinh(\lambda - \mu)\sinh (\lambda + \mu)}
\left( \sinh (2\lambda) \ F (\mu) -  \frac{\sinh (\xi - \lambda) }{\sinh(\xi - \mu)} \sinh (2\mu) \ F (\lambda) \right) \notag \\
&+ \frac{ 2\psi \sinh(2\lambda)}{\kappa \sinh(\lambda + \mu) \sinh (\xi + \lambda )} \ H (\mu) 
- \frac{\psi ^2 \sinh ^2 (2\mu)}{\kappa ^2 \sinh (\lambda + \mu ) \sinh (\xi - \mu )  \sinh (\xi + \mu )} \ E (\lambda) ,  \\[1ex]
\label{com-FF}
\left[ F (\lambda) , F (\mu) \right] &= \frac{ 2 \psi}{\kappa} \coth (\lambda + \mu) \left( \frac{\sinh (2\lambda)}{\sinh( \xi - \lambda)} \ F (\mu) - \frac{\sinh (2\mu)}{\sinh (\xi - \mu)} \ F (\lambda) \right)  \notag \\
&- \frac{2 \psi ^2}{\kappa ^2 \sinh(\lambda + \mu)} \left( \frac{\sinh ^2 ( 2\lambda)}{\sinh ( \xi - \lambda) \sinh ( \xi + \lambda)} \ H (\mu) -  \frac{\sinh ^2 (2\mu)}{\sinh (\xi - \mu)\sinh (\xi + \mu)} \ H (\lambda) \right) .
\end{align}

The Gaudin Hamiltonians with the boundary terms are obtained from
the residues of the generating function \eqref{eq: b-tau} 
%\begin{equation}
%\label{open-tau} 
%\tau (\lambda) =  \mathrm{tr}_0 \, \mathcal{L}_0 ^2(\lambda) 
%\end{equation}
 at  poles $\lambda = \pm\alpha_m$ :
\begin{equation}
\label{res-Ham}
\underset {\lambda = \alpha_m} {\mbox{Res}} \tau (\lambda) \ = \  4 \, H_m \qquad
\text{and} \qquad
\underset {\lambda = - \alpha_m} {\mbox{Res}} \tau (\lambda) \ =\ (-4) \,  H_m
\end{equation}
where
\begin{align}
\label{eq: open-Ham-a}
H_m &= \sum _{n \neq m}^N \left( \coth (\alpha _m - \alpha _n)  \ S^3_m S^3_n + \frac{S^+_m S^-_n + S^-_m S^+_n}{2 \sinh (\alpha _m - \alpha _n)} \right) + \sum _{n=1}^N \coth (\alpha _m + \alpha _n) \
\frac{S^3_m S^3_n + S^3_n S^3_m}{2} \notag \\[1ex]
&+ \frac{\psi}{\kappa} \ \frac{\sinh (2 \alpha _m)}{\sinh(\xi + \alpha_m)}  \sum _{n=1}^N \frac{S^3_m S^+_n + S^+_n S^3_m}{2 \sinh (\alpha _m + \alpha _n)} + \frac{\sinh (\xi - \alpha_m) }{2 \sinh (\xi + \alpha_m)} \sum _{n=1}^N \frac{S^-_m S^+_n + S^+_n S^-_m}{2 \sinh (\alpha _m + \alpha _n)}
\notag \\[1ex]
&- \frac{\psi}{\kappa} \ \frac{\sinh (2 \alpha _m)}{\sinh(\xi - \alpha_m)}  \sum _{n=1}^N \coth (\alpha _m + \alpha _n) \ \frac{S^+_m S^3_n + S^3_n S^+_m}{2} + \frac{\sinh (\xi + \alpha_m) }{2 \sinh (\xi - \alpha_m)} \sum _{n=1}^N \frac{S^+_m S^-_n + S^-_n S^+_m}{2 \sinh (\alpha _m + \alpha _n)}
\notag \\[1ex]
&- \frac{\psi ^2}{\kappa ^2} \ \frac{\sinh ^2 (2 \alpha _m )}{2 \sinh (\xi - \alpha _m) \sinh (\xi + \alpha _m)}
\sum _{n=1}^N \frac{S^+_m S^+_n + S^+_n S^+_m}{2 \sinh (\alpha _m + \alpha _n)} .
\end{align} 
The generating function of the Gaudin Hamiltonians \eqref{eq: b-tau} in terms of the entries of the Lax matrix is given by
\begin{equation}
\label{eq: open-t} 
\tau (\lambda) =  \mathrm{tr}_0 \, \mathcal{L}_0 ^2(\lambda) = 2 H^2(\lambda) + 2 F (\lambda) E (\lambda) + \left[ E (\lambda) , F (\lambda) \right] .
\end{equation} 
From \eqref{eq: com-EF} we have that the last term is
\begin{equation}
\left[ E (\lambda) , F (\lambda) \right] = 2 \frac{\cosh(2\xi) \cosh (2\lambda) -1}{\sinh (2\lambda)\sinh (\xi + \lambda)
\sinh(\xi - \lambda)} \ H(\lambda) - 2 H^{\prime}(\lambda) + \frac{2\psi \cosh(2\lambda)}{\kappa \sinh (\xi - \lambda)} \ E (\lambda) ,
\end{equation}
and therefore the final expression is
\begin{equation}
\label{eq: open-ta} 
\tau (\lambda) = 2 \left( H^2(\lambda) + \frac{\cosh(2\xi) \cosh (2\lambda) -1}{\sinh (2\lambda)\sinh (\xi + \lambda) \sinh(\xi - \lambda)} \ H(\lambda) -  H^{\prime}(\lambda)  \right) + \left( 2 F (\lambda) + \frac{2\psi \cosh(2\lambda)}{\kappa \sinh (\xi - \lambda)}  \right) E (\lambda) .
\end{equation}

Our next step is to introduce the new generators $e(\lambda) , h(\lambda)$ and $f(\lambda)$ as the following linear combinations of the original ones
\begin{align}
\label{eq: new-basis-e}
e(\lambda) &= \frac{\sinh(\xi + \lambda)}{\sinh (2\lambda)} E(\lambda) = \sum_{m=1}^N \frac{\sinh(\xi+\alpha_m)  \ S^+_m}{\sinh(\lambda -\alpha_m)\sinh(\lambda +\alpha_m)} , \\[1ex]
\label{eq: new-basis-h}
h(\lambda) &= \frac{1}{\sinh (2 \lambda)} \left( H(\lambda) - \frac{\psi \sinh (\lambda)}{k \sinh (\xi)} \ E(\lambda) \right) = \sum_{m=1}^N \frac{S^3_m - \displaystyle{\frac{\psi \sinh(\alpha_m)}{\kappa \sinh(\xi)}} \ S^+_m}{\sinh(\lambda -\alpha_m)\sinh(\lambda +\alpha_m)} ,  \\[1ex]
\label{eq: new-basis-f}
f(\lambda) &= \frac{1}{\sinh (2\lambda)} \left( \sinh (\xi - \lambda) F(\lambda) +  \frac{\psi}{\kappa} \sinh (2\lambda) H (\lambda) \right) = \sum_{m=1}^N \frac{\sinh(\xi - \alpha_m)  \ S^-_m + \displaystyle{\frac{\psi}{\kappa}} \sinh (2\alpha_m) \ S^3_m}{\sinh(\lambda -\alpha_m)\sinh(\lambda + \alpha_m)} .
\end{align}
The key observation is that in the new basis  we have
\begin{equation}
\label{eq: 0-comm}
\left[ e(\lambda) , e(\mu) \right] = \left[ h(\lambda) , h(\mu) \right] = \left[ f(\lambda) , f(\mu) \right] = 0 .
\end{equation}
Therefore there are only three nontrivial commutation relations
\begin{align}
\label{eq: h-e}
\left[ h(\lambda) , e(\mu) \right] &= \frac{1}{\sinh(\lambda - \mu)\sinh(\lambda + \mu)} \left( e (\mu) - e (\lambda) \right) ,  \\[1ex]
\label{eq: h-f}
\left[ h(\lambda) , f(\mu) \right]  &=  \frac{- 1}{\sinh(\lambda - \mu)\sinh(\lambda + \mu)} 
\left( f (\mu) - f (\lambda) \right)  + \frac{2 \psi \coth(\xi)}{\kappa \sinh(\lambda - \mu)\sinh(\lambda + \mu)} \times \notag \\
&\times  \left( \sinh ^2 (\mu) \, h (\mu) - \sinh ^2(\lambda) \, h (\lambda) \right) + \frac{ 2 \psi ^2}{\kappa ^2 \sinh(\lambda - \mu)\sinh(\lambda + \mu) \sinh ^2(\xi)} \times \notag \\
&\times \left(  \sinh ^2 (\mu) \, e (\mu) - \sinh ^2(\lambda) \, e(\lambda) \right) ,  \\[1ex]
\label{eq: e-f}
\left[ e(\lambda) , f(\mu) \right]  &=  \frac{- 2 \psi \coth(\xi)}{\kappa \sinh(\lambda - \mu)\sinh(\lambda + \mu)}\left( \sinh ^2(\mu) \, e (\mu) - \sinh ^2(\lambda) \, e (\lambda) \right) 
+ \frac{2}{\sinh(\lambda - \mu)\sinh(\lambda + \mu)} \times \notag \\
& \times \left( \sinh(\xi - \mu)\sinh(\xi + \mu) \, h (\mu) - \sinh(\xi - \lambda)\sinh(\xi + \lambda) \, h (\lambda) \right) .
\end{align}

Our aim is to implement the algebraic Bethe ansatz  based on the Lie algebra \eqref{eq: 0-comm} - \eqref{eq: e-f}. To this end we need to obtain the expression for the generating function $\tau (\lambda)$ in terms of the generators $e(\lambda) , h(\lambda)$ and $f(\lambda)$. The first step is to invert the relations \eqref{eq: new-basis-e} - \eqref{eq: new-basis-f}
\begin{align}
\label{eq: inv-E}
E(\lambda) &= \frac{\sinh (2\lambda)}{\sinh(\xi + \lambda)} \ e(\lambda) ,  \\[1ex]
\label{eq: inv-H}
H(\lambda) &= \sinh (2\lambda) \left( h(\lambda) + \frac{\psi \sinh (\lambda)}{\kappa \sinh(\xi) \sinh(\xi + \lambda)} \ e (\lambda) \right) ,  \\[1ex]
\label{eq: inv-F}
F(\lambda) &= \frac{\sinh(2\lambda)}{\sinh (\xi - \lambda)} \left( f(\lambda) - \frac{\psi \sinh (2\lambda)}{\kappa} \  h (\lambda) - \frac{\psi ^2 \sinh (\lambda) \sinh(2\lambda)}{\kappa ^2 \sinh(\xi) \sinh (\xi + \lambda)} \ e(\lambda) \right) .
\end{align}

In particular, we have
\begin{align}
\label{eq: H2}
H^2(\lambda) &= \sinh ^2  (2\lambda) \left( h ^2 (\lambda) + \frac{\psi \sinh (\lambda)}{\kappa \sinh(\xi) \sinh(\xi + \lambda)} \left( 2 h (\lambda) e (\lambda) - \left[  h (\lambda) , e (\lambda)\right] \right) \right. \notag \\
&\left. +\frac{\psi ^2\sinh ^2(\lambda)}{\kappa^2 \sinh ^2 (\xi) \sinh ^2(\xi + \lambda)}  \ e ^2(\lambda)  \right)  \notag \\[1ex]
&=  \sinh ^2  (2\lambda) \left( h ^2 (\lambda) + \frac{\psi \sinh (\lambda)}{\kappa \sinh(\xi) \sinh(\xi + \lambda)} \left( 2 h (\lambda) e (\lambda) + \frac {e^{\prime} (\lambda)}{\sinh (2\lambda)}\right) \right. \notag \\
&\left. +\frac{\psi ^2\sinh ^2(\lambda)}{\kappa^2 \sinh ^2 (\xi) \sinh ^2(\xi + \lambda)}  \ e ^2(\lambda)  \right) .   
\end{align}
Substituting \eqref{eq: inv-E} -- \eqref{eq: H2} into \eqref{eq: open-ta} we obtain the desired expression for the generating function
\begin{equation}
\label{eq: open-tau} 
\begin{split}
%\!\!\!\!\!\!\!\!\!\!\!\!\!\!\!\!\!\!\!\!\!\!\!\!
\tau (\lambda) &= 2 \sinh ^2(2 \lambda) \left( h ^2(\lambda) + \frac{h (\lambda)}{\sinh (\xi + \lambda) \sinh (\xi - \lambda)}- \frac{ h ^{\prime}(\lambda)}{\sinh (2\lambda)} \right) + \frac{2 \sinh ^2(2 \lambda)}{\sinh (\xi + \lambda) \sinh (\xi - \lambda)} \times
\\[1ex]
& \times \left( f (\lambda) - 2 \frac{\psi}{\kappa} \coth (\xi) \sinh ^2(\lambda) \ h (\lambda) - 
\frac{\psi ^2 \sinh ^2(\lambda)}{\kappa ^2 \sinh ^2(\xi)} \ e (\lambda) + \frac{\psi}{\kappa} \coth (\xi) \right) e (\lambda) .
\end{split}
\end{equation} 

In the Hilbert space $\mathcal{H}$ \eqref{eq: H-space},  in every $V_ m = \mathbb{C}^{2s+1}$ there exists a vector $\omega_m \in V_ m$ such that
\begin{equation}
\label{eq: S-on-om}
S^3_m \omega _m = s_m \omega _m  \quad \text{and}  \quad S^+_m \omega _m = 0 .
\end{equation}
We define a vector $\Omega _+$ to be
\begin{equation}
\label{eq: Omega+}
\Omega _+ = \omega _1 \otimes \cdots \otimes \omega _N \in \mathcal{H}.
\end{equation} 
From the definitions above, the formulas \eqref{gen-E} - \eqref{gen-H} and  \eqref{eq: new-basis-e} -  \eqref{eq: new-basis-f} it is straightforward to obtain the action of the generators $e(\lambda)$  and $h(\lambda)$ on the vector $\Omega _+$
\begin{equation}
\label{eq: action-Om}
e(\lambda) \Omega _+ = 0 \ \ \text{and} \ \
h(\lambda) \Omega _+ = \rho (\lambda) \Omega _+, \ \ \text{with} \ \
\rho (\lambda) = \sum _{m=1}^N \frac{s_m}{\sinh (\lambda + \alpha _m) \sinh (\lambda - \alpha _m)}.
\end{equation}
An important initial observation in the implementation of the algebraic Bethe ansatz is that the vector  $\Omega _+$ \eqref{eq: Omega+} is an eigenvector of the generating function $\tau (\lambda) $. To show this we use \eqref{eq: action-Om}
\begin{equation}
\label{eq: tau-Om}
\tau (\lambda) \Omega _+ = \chi _0 (\lambda) \Omega _+ = 2 \sinh ^2 (2\lambda)  \left( \rho ^2 (\lambda) + \frac{\rho (\lambda)}{\sinh (\xi + \lambda) \sinh (\xi - \lambda)}   - \frac{\rho^{\prime} (\lambda)}{\sinh (2\lambda)} \right) \Omega _+ ,
\end{equation}
and by substituting the function $\rho (\lambda)$ \eqref{eq: action-Om} the eigenvalue $\chi _0 (\lambda)$ can be expressed as
\begin{equation}
\label{eq: chi-0}
\begin{split}
\chi _0 (\lambda) &= 2 \sinh ^2 (2\lambda) \left( \sum _{m=1}^N \frac{s_m (s_m+1)}{\sinh ^2(\lambda + \alpha _m) \sinh ^2(\lambda - \alpha _m)} + \sum _{m=1}^N \frac{s_m}{\sinh (\lambda + \alpha _m) \sinh (\lambda - \alpha _m)} \times \right. \\
&\left. \times \left( \frac{1}{\sinh (\xi + \lambda) \sinh (\xi - \lambda)} +
\sum _{n > m}^N \frac{2s_n}{\sinh (\lambda + \alpha _n) \sinh (\lambda - \alpha _n)} \right) \right) .
\end{split}
\end{equation}

An essential step in the algebraic Bethe ansatz is the definition of the corresponding Bethe vectors. In this case, they are symmetric functions of their arguments and are such that the off-shell action of the generating function of the Gaudin Hamiltonians is as simple as possible. With this aim we proceed to show that the Bethe vector $\varphi _1(\mu)$ has the form
\begin{equation}
\label{eq: phi-1}
\varphi _1(\mu) = \left( f(\mu) + c_1^{(1)}(\mu) \right) \Omega _+,
\end{equation}
where $c_1(\mu)$ is given by
\begin{equation}
\label{eq: c-1}
c_1^{(1)}(\mu) = \frac{\psi}{\kappa} \left( 1 + \left( e^{-2\xi} - \cosh(2\mu) \right) \ \rho(\mu) \right) .
\end{equation} 
Evidently, the action of the generating function of the Gaudin Hamiltonians reads
\begin{equation}
\label{eq: tau-p1}
\tau (\lambda) \varphi _1(\mu) = \left[ \tau (\lambda) , f(\mu) \right] \Omega _+ + \chi _0 (\lambda) \varphi _1(\mu) .
 \end{equation}
A straightforward calculation shows that  the commutator in the first term of \eqref{eq: tau-p1} is given by
\begin{align}
\label{eq: com-t-f}
\left[ \tau (\lambda) , f(\mu) \right] \Omega _+ &= - \frac{2 \sinh ^2 (2 \lambda )}{\sinh (\lambda + \mu) \sinh (\lambda - \mu)} \left( 2 \rho (\lambda) + \frac{1}{\sinh (\xi + \lambda)\sinh (\xi -\lambda )} \right) \varphi _1(\mu) \notag \\[1ex]
 &+ \frac{2 \sinh ^2 (2 \lambda )}{\sinh (\lambda + \mu) \sinh (\lambda - \mu)} \frac{\sinh (\xi+\mu)\sinh (\xi -\mu)}{\sinh (\xi + \lambda)\sinh (\xi -\lambda)} \times \notag \\[1ex]
&\times \left(  2 \rho (\mu) + \frac{1}{\sinh (\xi+\mu)\sinh (\xi -\mu)} \right) \varphi _1(\lambda) .
\end{align}
Therefore the action of the generating function $\tau (\lambda)$ on $\varphi _1(\mu)$ is given by
\begin{equation}
\label{eq: tau-phi-1}
\begin{split}
\tau (\lambda) \varphi _1(\mu) &= \chi _1 (\lambda, \mu) \varphi _1(\mu) + \frac{2 \sinh ^2 (2 \lambda)}{\sinh (\lambda + \mu) \sinh (\lambda - \mu)} \frac{\sinh (\xi+\mu)\sinh (\xi -\mu)}{\sinh (\xi + \lambda)\sinh (\xi -\lambda)} \times \\
&\times \left(  2 \rho (\mu) + \frac{1}{\sinh (\xi+\mu)\sinh (\xi -\mu)} \right) \varphi _1(\lambda) .
\end{split}
\end{equation}
with
\begin{equation}
\label{eq: chi-1}
\chi _1 (\lambda, \mu) = \chi _0 (\lambda) - \frac{2 \sinh ^2 (2 \lambda )}{\sinh (\lambda + \mu) \sinh (\lambda - \mu)} \left( 2 \rho (\lambda) + \frac{1}{\sinh (\xi + \lambda)\sinh (\xi -\lambda )} \right).
\end{equation}
The unwanted term in \eqref{eq: tau-phi-1} vanishes when the following Bethe equation is imposed on the parameter $\mu$,
\begin{equation}
\label{eq: BE-1}
2 \rho (\mu) + \frac{1}{\sinh (\xi+\mu)\sinh (\xi -\mu)} = 0 .
\end{equation}
Thus we have shown that $\varphi _1(\mu)$ \eqref{eq: phi-1} is the desired Bethe vector of the generating function $\tau (\lambda)$ corresponding to the eigenvalue $\chi _1 (\lambda, \mu)$ \eqref{eq: chi-1}.

We seek the Bethe vector $\varphi _2 (\mu _1, \mu _2)$ as the following symmetric function
\begin{equation}
\label{eq: phi-2}
\varphi _2(\mu _1, \mu _2) = f(\mu _1) f(\mu _2) \Omega _+ 
+ c_2^{(1)}(\mu _2 ; \mu _1) f(\mu _1) \Omega _+
+ c_2^{(1)}(\mu _1 ; \mu _2) f(\mu _2) \Omega _+
+ c_2^{(2)}(\mu _1, \mu _2) \Omega _+,
\end{equation}
where the scalar coefficients $c_2^{(1)}(\mu _1 ; \mu _2)$  and $c_2^{(2)}(\mu _1, \mu _2)$ are
\begin{align}
\label{eq: c2-1}
c_2^{(1)}(\mu _1 ; \mu _2)  &= \frac{\psi}{\kappa} \left( 1 + \left( e^{-2\xi} - \cosh( 2 \mu_1 )\right) \left( \rho(\mu _1) - \frac{1}{\sinh (\mu_1 - \mu_2) \sinh (\mu_1 + \mu_2)} \right) \right) , \\[1ex]
\label{eq: c2-2aux}
c_2^{(2)}(\mu _1, \mu _2)  &= \frac{\psi ^2}{\kappa ^2} \left( 3 + \left( e^{-2\xi} - \cosh( 2 \mu_1 ) \right)  \left( e^{-2\xi} - \cosh( 2 \mu_2 ) \right)   \rho(\mu_1) \rho(\mu_2) +
\right. \notag \\
&+ \frac{2 e^{-4\xi} + 2 e^{-2\xi} \left( \cosh(2\mu_1) - 3 \cosh (2\mu_2) \right) - \left( 3 + \cosh (4\mu_1) - 6 \cosh (2\mu_1) \cosh (2\mu_2) \right)}{4 \sinh (\mu_1 - \mu_2) \sinh (\mu_1 + \mu_2)} \rho(\mu_1)  \notag \\
&\left.  + \frac{2 e^{-4\xi} + 2 e^{-2\xi} \left( \cosh(2\mu_2) - 3 \cosh (2\mu_1) \right) - \left( 3 + \cosh (4\mu_2) - 6 \cosh (2\mu_2) \cosh (2\mu_1) \right)}{4 \sinh (\mu_2 - \mu_1) \sinh (\mu_2 + \mu_1)} \rho(\mu_2) \right) .
\end{align}
One way to obtain the action of $\tau (\lambda)$ on $\varphi _2(\mu _1, \mu _2)$ is to write
\begin{equation}
\label{eq: calc-tau-phi-2}
\begin{split}
\tau (\lambda) \varphi _2 (\mu _1, \mu _2) &= \left[ \left[ \tau (\lambda) , f(\mu _1) \right] , f(\mu _2) \right] \Omega _+  + \left( f(\mu _2) + c_2^{(1)}(\mu _2 ; \mu _1) \right) \left[ \tau (\lambda) , f(\mu _1) \right] \Omega _+ \\[1ex]
& + \left( f(\mu _1) + c_2^{(1)}(\mu _1 ; \mu _2) \right) \left[ \tau (\lambda) , f(\mu _2) \right] \Omega _+
+ \chi _0 (\lambda) \varphi _2 (\mu _1, \mu _2) .
\end{split}
\end{equation}
Then, we substitute \eqref{eq: com-t-f} in the second and third term above and use the relations
\begin{align}
\label{eq: rel-phi-2-mu}
&\left( f(\mu _1) + c_2^{(1)}(\mu _1 ; \mu _2) \right)  \varphi _1 (\mu _2) = \varphi _2 (\mu _1, \mu _2) - \frac{\psi}{\kappa} \ \frac{e^{-2\xi} - \cosh (2 \mu_2)}{\sinh (\mu_1 - \mu_2) \sinh (\mu_1 + \mu_2)} \varphi _1 (\mu _1) \notag \\[1ex]
 &- \left( c_2^{(2)}(\mu _1, \mu _2) - c_1^{(1)}(\mu _1) c_1^{(1)}(\mu _2) + \frac{\psi}{\kappa} \ \frac{ \left( e^{-2\xi} - \cosh (2 \mu_1) \right) c_1^{(1)}(\mu _2) - \left( e^{-2\xi} - \cosh (2 \mu_2) \right)c_1^{(1)}(\mu _1)}{\sinh (\mu_1 - \mu_2) \sinh (\mu_1 + \mu_2)} \right) \Omega _+ , \\
\label{eq: rel-phi-2-l}
&\left( f(\mu _1) + c_2^{(1)}(\mu _1 ; \mu _2) \right)  \varphi _1 (\lambda) = \varphi _2 (\mu _1, \lambda) - \frac{\psi}{\kappa}\ \frac{e^{-2\xi} - \cosh (2 \lambda)}{\sinh (\lambda - \mu_1) \sinh (\lambda + \mu_1)} \varphi _1 (\mu _1) \notag \\[1ex] 
&+ \left( c_2^{(1)}(\mu _1; \mu _2) - c_2^{(1)}(\mu _1; \lambda) \right) \varphi _1 (\lambda) \notag \\[1ex] 
&- \left( c_2^{(2)}(\mu _1, \lambda) - c_1^{(1)}(\mu _1) c_1^{(1)}(\lambda) + \frac{\psi}{\kappa} \ \frac{\left( e^{-2\xi} - \cosh (2 \lambda) \right) c_1^{(1)}(\mu _1) - \left( e^{-2\xi} - \cosh (2 \mu_1) \right) c_1^{(1)}(\lambda)}{\sinh (\lambda - \mu_1) \sinh (\lambda + \mu_1)} \right) \Omega _+ ,
\end{align}
which follow from the definition \eqref{eq: phi-2}. After expressing appropriately the first term on the right-hand side of \eqref{eq: calc-tau-phi-2} and using twice the expression for the action of the commutator of $\tau (\lambda)$ with the generator $f(\lambda)$ on the vector $\Omega _+$ \eqref{eq: com-t-f} as well as the identities \eqref{eq: rel-phi-2-mu} and \eqref{eq: rel-phi-2-l}, a
straightforward calculation shows that the off-shell action of the generating function $\tau (\lambda)$ on $\varphi _2 (\mu _1, \mu _2)$ is given by
\begin{equation}
\label{eq: tau-phi-2}
\begin{split}
\tau (\lambda) \varphi _2 (\mu _1, \mu _2) &= \chi _2 (\lambda, \mu _1, \mu _2 ) \varphi _2 (\mu _1, \mu _2) 
+ \sum _{i=1}^2 \frac{2 \sinh ^2 (2 \lambda)}{\sinh (\lambda + \mu _i) \sinh (\lambda - \mu _i)} \frac{\sinh (\xi +\mu _i) \sinh (\xi - \mu _i)}{\sinh (\xi + \lambda)\sinh (\xi -\lambda)} \times  \\
& \times \left(  2 \rho (\mu _i) + \frac{1}{\sinh (\xi + \mu _i)\sinh (\xi -\mu _i)} - \frac{2}{\sinh (\mu_i + \mu _{3-i}) \sinh (\mu_i - \mu _{3-i})} 
\right) \varphi _2 (\lambda, \mu _{3-i}) , 
\end{split}
\end{equation}
with the eigenvalue
\begin{equation}
\label{eq: chi-2}
\begin{split}
&\chi _2 (\lambda, \mu _1, \mu _2) = \chi _0 (\lambda) - \sum _{i=1}^2 \frac{2 \sinh ^2 (2 \lambda)}{\sinh (\lambda + \mu _i) \sinh (\lambda - \mu _i)} \times \\
&\times \left(2 \rho (\lambda) + \frac{1}{\sinh (\xi + \lambda)\sinh (\xi -\lambda )} - \frac{1}{\sinh ( \lambda + \mu _{3-i} ) \sinh (\lambda - \mu _{3-i} )} \right).   
\end{split}
\end{equation}
The two unwanted terms in the action above \eqref{eq: tau-phi-2} vanish when the Bethe equations are imposed on the parameters $\mu _1$ and $\mu _2$,
\begin{equation}
\label{eq: BE-2}
2 \rho (\mu _i) + \frac{1}{\sinh (\xi + \mu _i)\sinh (\xi -\mu _i)} - \frac{2}{\sinh (\mu_i + \mu _{3-i}) \sinh (\mu_i - \mu _{3-i})}  = 0 ,
\end{equation}
with $i=1,2$. Therefore $\varphi _2 (\mu _1, \mu _2)$ is the Bethe vector of the generating function of the Gaudin Hamiltonians with the eigenvalue $\chi _2 (\lambda, \mu _1, \mu _2)$.

As a symmetric function of its arguments the Bethe vector $\varphi _3 (\mu _1, \mu _2, \mu _3)$ is given explicitly in the Appendix B. A lengthy but straightforward calculation based on appropriate generalisation of \eqref{eq: calc-tau-phi-2} - \eqref{eq: rel-phi-2-l} shows that the action of the generating function $\tau (\lambda)$ on $\varphi _3 (\mu _1, \mu _2, \mu _3)$ is given by
\begin{equation}
\label{eq: tau-phi-3}
\begin{split}
&\tau (\lambda) \varphi _3 (\mu _1, \mu _2, \mu _3) = \chi _3 (\lambda,\mu _1, \mu _2 , \mu _3) 
\varphi _3 (\mu _1, \mu _2, \mu _3) \\[1ex]
&+ \sum _{i=1}^3 \frac{2 \sinh ^2 (2 \lambda)}{\sinh (\lambda + \mu _i) \sinh (\lambda - \mu _i)} 
\frac{\sinh (\xi +\mu _i) \sinh (\xi - \mu _i)}{\sinh (\xi + \lambda)\sinh (\xi -\lambda)} \times \\
& \times %\frac{\sinh (\xi +\mu _i) \sinh (\xi - \mu _i)}{\sinh (\xi + \lambda)\sinh (\xi -\lambda)}
\left( 2 \rho (\mu _i) + \frac{1}{\sinh (\xi + \mu _i)\sinh (\xi -\mu _i)} - \sum _{j\neq i}^3 \frac{2}{\sinh (\mu_i + \mu _{j}) \sinh (\mu_i - \mu _{j})} \right) \varphi _3 (\lambda, \{ \mu _j \} _{j\neq i} ) , 
\end{split}
\end{equation}
where the eigenvalue is
\begin{equation}
\label{eq: chi-3}
\begin{split}
&\chi _3 (\lambda,\mu _1, \mu _2, \mu _3) = \chi _0 (\lambda) - \sum _{i=1}^3  \frac{2 \sinh ^2 (2 \lambda)}{\sinh (\lambda + \mu _i) \sinh (\lambda - \mu _i)} \times \\
&\times \left(2 \rho (\lambda) + \frac{1}{\sinh (\xi + \lambda)\sinh (\xi -\lambda )} - \sum _{j\neq i}^3 \frac{1}{\sinh ( \lambda + \mu _{j} ) \sinh (\lambda - \mu _{j} )} \right).   
\end{split}
\end{equation}
The three unwanted terms in \eqref{eq: tau-phi-3} vanish when the Bethe equation are imposed on the parameters $\mu _i$,
\begin{equation}
\label{eq: BE-3}
%\beta _3(\mu _i; \{ \mu_j \} _{j\neq i}) = 
2 \rho (\mu _i) + \frac{1}{\sinh (\xi + \mu _i)\sinh (\xi -\mu _i)}  - \sum _{j\neq i}^3 \frac{2}{\sinh (\mu_i + \mu _{j}) \sinh (\mu_i - \mu _{j})}  = 0 ,
\end{equation}
with $i=1,2,3$.

Although it would be natural to continue this approach and present here the Bethe vector $\varphi _4 ( \mu_1 ,  \mu_2, \mu_3, \mu_4 )$ as a symmetric function of its arguments, it turns out that the expressions for the coefficients functions $c ^{(i)}_4 (\mu_1, \dots , \mu_i ; \mu_{i+1},  \ldots , \mu_4)$, $i=1, 2, 3, 4$, are cumbersome, not admitting any compact form. For this reason we have decided not present them here. Instead we define the family of operators
\begin{equation}
\label{eq: C_K}
\begin{split}
\mathcal{C}_K(\mu) &= f(\mu ) + \frac{\psi}{\kappa} \left( \left( 2K-1 \right) + \left( e^{-2\xi} - \cosh (2\mu) \right) h(\mu) \right) + \frac{\psi ^2}{\kappa ^2} \frac{e^{-\xi}}{2\sinh(\xi)} \times \\
&\times \left( e^{-2\xi} + 1 - 2 \cosh (2\mu) \right) e(\mu) ,
\end{split}
\end{equation}
for any natural number $K$. A direct calculation shows that the Bethe vectors \eqref{eq: phi-1}, \eqref{eq: phi-2} and \eqref{eq: phi-3} can be expressed as
\begin{equation}
\label{eq: phi1,2,3-c}
\varphi _1 (\mu ) = \mathcal{C}_1(\mu) \Omega _+ ,\quad 
\varphi _2 (\mu _1, \mu_2) = \mathcal{C}_1(\mu _1) \mathcal{C}_2(\mu_2) \Omega _+ \quad \text{and} \quad
\varphi _3 (\mu _1, \mu_2,\mu_3) = \mathcal{C}_1(\mu _1) \mathcal{C}_2(\mu_2) \mathcal{C}_3(\mu_3)\Omega _+ .
\end{equation}
Although in general the operators $\mathcal{C}_K(\mu)$ \eqref{eq: C_K} do not commute, it is effortless to confirm that the Bethe vector $\varphi _2 (\mu _1, \mu_2)$ is a symmetric function
\begin{equation}
\label{eq: phi2-sym}
\varphi _2 (\mu _1, \mu_2) = \mathcal{C}_1(\mu _1) \mathcal{C}_2(\mu_2) \Omega _+ = \mathcal{C}_1(\mu _2) \mathcal{C}_2(\mu_1) \Omega _+ = \varphi _2 (\mu _2, \mu_1) . 
\end{equation}
Analogously, it is straightforward to check that the Bethe vector $\varphi _3 (\mu _1, \mu_2,\mu_3)$ is a symmetric function of its arguments
\begin{equation}
\label{eq: phi3-sym}
\varphi _3 (\mu _1, \mu_2,\mu_3) = \mathcal{C}_1(\mu _1) \mathcal{C}_2(\mu_2) \mathcal{C}_3(\mu_3) \Omega _+ = \mathcal{C}_1(\mu _2) \mathcal{C}_2(\mu_1) \mathcal{C}_3(\mu_3) \Omega _+ = \varphi _3 (\mu _2, \mu_1, \mu_3) , 
\end{equation}
etc. Moreover using the formulae above \eqref{eq: phi1,2,3-c} for the Bethe vectors it is somewhat simpler to calculate the off-shell action of the generating function. Evidently,
$$
\left[ \tau (\lambda), \mathcal{C}_1(\mu ) \right] \Omega _+ = \left[ \tau (\lambda), f(\mu ) \right] \Omega _+ ,
$$
and consequently, the action \eqref{eq: tau-phi-1} of the generating function $\tau (\lambda)$ on the Bethe vector $\varphi _1 (\mu )$ follows directly from \eqref{eq: com-t-f}. In order to show \eqref{eq: tau-phi-2}, we calculate
\begin{equation}
\label{eq: tau-cc}
\begin{split}
&\left[ \left[ \tau (\lambda) , \mathcal{C}_1(\mu _1) \right] , \mathcal{C}_2(\mu _2) \right] \Omega _+ =
\frac{4\sinh ^2(2\lambda)}{\sinh (\lambda + \mu_1) \sinh (\lambda - \mu_1)\sinh (\lambda + \mu_2) \sinh (\lambda - \mu_2)} \varphi _2 (\mu_1,\mu_2) \\[1ex]
& - \sum_{i=1}^2 \frac{2\sinh ^2(2\lambda)}{\sinh (\lambda + \mu_i) \sinh (\lambda - \mu_i)} \ \frac{\sinh (\xi + \mu_i) \sinh (\xi - \mu_i)}{\sinh (\xi + \lambda) \sinh (\xi -\lambda)} \ \frac{2}{\sinh (\mu_i + \mu_{3-i}) \sinh (\mu_i - \mu_{3-i})} \varphi _2 (\lambda,\mu_{3-i}) \\[1ex]
&+ \frac{\psi}{\kappa} \frac{4 \sinh ^2(2\lambda)}{\sinh (\lambda + \mu_1) \sinh (\lambda - \mu_1)} 
\left(2 \rho (\lambda) + \frac{1}{\sinh (\xi + \lambda)\sinh (\xi -\lambda )} \right) \left( \varphi _1 (\mu_1) - \varphi _1 (\mu_2)\right) \\[1ex]
&- \frac{\psi}{\kappa} \frac{4 \sinh ^2(2\lambda)}{\sinh (\lambda + \mu_1) \sinh (\lambda - \mu_1)} 
\frac{\sinh (\xi + \mu_1) \sinh (\xi - \mu_1)}{\sinh (\xi + \lambda) \sinh (\xi -\lambda)}
\left(2 \rho (\mu_1) + \frac{1}{\sinh (\xi + \mu_1)\sinh (\xi -\mu_1 )} \right) \times \\[1ex]
&\times \left( \varphi _1 (\lambda) - \varphi _1 (\mu_2)\right) \\[1ex]
&- \frac{\psi}{\kappa} \frac{4 \sinh ^2(2\lambda)}{\sinh (\lambda + \mu_2) \sinh (\lambda - \mu_2)} 
\left(2 \rho (\lambda) + \frac{1}{\sinh (\xi + \lambda)\sinh (\xi -\lambda )} \right) \varphi _1 (\mu_1)  \\[1ex]
&+ \frac{\psi}{\kappa} \frac{4 \sinh ^2(2\lambda)}{\sinh (\lambda + \mu_2) \sinh (\lambda - \mu_2)} 
\frac{\sinh (\xi + \mu_2) \sinh (\xi - \mu_2)}{\sinh (\xi + \lambda) \sinh (\xi -\lambda)}
\left(2 \rho (\mu_2) + \frac{1}{\sinh (\xi + \mu_2)\sinh (\xi -\mu_2)} \right) \varphi _1 (\mu_1) 
%\\[1ex]
\end{split}
\end{equation}

and use \eqref{eq: com-t-f} appropriately. Finally, the action \eqref{eq: tau-phi-3} of the generating function $\tau (\lambda)$ on the Bethe vector $\varphi _3 (\mu _1, \mu_2,\mu_3)$ can be obtained
by expressing $\left[ \left[ \left[ \tau (\lambda) , \mathcal{C}_1(\mu _1) \right] , \mathcal{C}_2(\mu _2) \right] ,  \mathcal{C}_3(\mu _3) \right] \Omega _+$ conveniently and using  \eqref{eq: tau-cc} and \eqref{eq: com-t-f} adequately.

Therefore we look for the Bethe vector $\varphi _4 (\mu _1, \ldots,\mu_4)$ in the form
\begin{equation}
\label{eq: phi-4}
\varphi _4 (\mu _1, \ldots,\mu_4) = \mathcal{C}_1(\mu _1) \mathcal{C}_2(\mu_2) \mathcal{C}_3(\mu_3) \mathcal{C}_4(\mu_4) \Omega _+ .
\end{equation}
With the aim of calculating the action of the generating function of the Gaudin Hamiltonians on the vector above we calculate $\left[ \left[ \left[ \left[ \tau (\lambda) , \mathcal{C}_1(\mu _1) \right] , \mathcal{C}_2(\mu _2) \right] ,  \mathcal{C}_3(\mu _3) \right] , \mathcal{C}_4(\mu_4) \right] \Omega _+$, expressing it appropriately as a linear combination of all the previous Bethe vectors. This formula is very long and cumbersome and for this reason, is not presented in the text. Using this result it is possible to obtain the desired off-shell action in the following form
\begin{equation}
\label{eq: tau-phi-4}
\begin{split}
&\tau (\lambda) \varphi _4 (\mu _1, \mu _2, \mu _3, \mu _4) = \chi _4 (\lambda,\mu _1, \mu _2 , \mu _3, \mu _4) \varphi _4 (\mu _1, \mu _2, \mu _3, \mu _4)  \\[1ex]
&+ \sum _{i=1}^4 \frac{2 \sinh ^2 (2 \lambda)}{\sinh (\lambda + \mu _i) \sinh (\lambda - \mu _i)} 
\frac{\sinh (\xi +\mu _i) \sinh (\xi - \mu _i)}{\sinh (\xi + \lambda)\sinh (\xi -\lambda)} \times \\
& \times %\frac{\sinh (\xi +\mu _i) \sinh (\xi - \mu _i)}{\sinh (\xi + \lambda)\sinh (\xi -\lambda)}
\left( 2 \rho (\mu _i) + \frac{1}{\sinh (\xi + \mu _i)\sinh (\xi -\mu _i)} - \sum _{j\neq i}^4 \frac{2}{\sinh (\mu_i + \mu _{j}) \sinh (\mu_i - \mu _{j})} \right) \varphi _4 (\lambda, \{ \mu _j \} _{j\neq i} ) , 
\end{split}
\end{equation}
where the eigenvalue is
\begin{equation}
\label{eq: chi-4}
\begin{split}
&\chi _4 (\lambda,\mu _1, \mu _2, \mu _3, \mu _4) = \chi _0 (\lambda) - \sum _{i=1}^4  \frac{2 \sinh ^2 (2 \lambda)}{\sinh (\lambda + \mu _i) \sinh (\lambda - \mu _i)} \times \\
&\times \left(2 \rho (\lambda) + \frac{1}{\sinh (\xi + \lambda)\sinh (\xi -\lambda )} - \sum _{j\neq i}^4 \frac{1}{\sinh ( \lambda + \mu _{j} ) \sinh (\lambda - \mu _{j} )} \right).   
\end{split}
\end{equation}
This result we have confirmed also by symbolic computing.  The four unwanted terms in \eqref{eq: tau-phi-4} vanish when the Bethe equation are imposed on the parameters $\mu _i$,
\begin{equation}
\label{eq: BE-4}
%\beta _3(\mu _i; \{ \mu_j \} _{j\neq i}) = 
2 \rho (\mu _i) + \frac{1}{\sinh (\xi + \mu _i)\sinh (\xi -\mu _i)}  - \sum _{j\neq i}^4 \frac{2}{\sinh (\mu_i + \mu _{j}) \sinh (\mu_i - \mu _{j})}  = 0 ,
\end{equation}
with $i=1,2,3,4$. 

We readily proceed to define $\varphi _M ( \mu_1 ,  \mu_2 ,  \dots ,  \mu_M )$, for an arbitrary positive integer $M$,
\begin{equation}
\label{eq: phi-M}
\varphi _M (\mu _1,  \mu_2 , \ldots,\mu_M) = \mathcal{C}_1(\mu _1) \mathcal{C}_2(\mu _2) \cdots \mathcal{C}_M(\mu_M) \Omega _+ ,
\end{equation}
and the operators $\mathcal{C}_K(\mu )$ are given in \eqref{eq: C_K}. Although the operators $\mathcal{C}_K(\mu )$ do not commute, the Bethe vector $\varphi_M ( \mu_1 ,  \mu_2 ,  \dots ,  \mu_M )$ is a symmetric function of its arguments, since these operators satisfy the following identity,
\begin{equation}
\label{C-commute}
\mathcal{C}_K(\mu) \mathcal{C}_{K+1}(\widetilde{\mu}) - \mathcal{C}_K(\widetilde{\mu}) \mathcal{C}_{K+1}(\mu) = 0,
\end{equation}
for $K=1, \ldots , M-1$. It is possible to check that the off-shell action of the generating function $\tau (\lambda)$ on the Bethe vector $\varphi _M ( \mu_1 ,  \mu_2 ,  \dots ,  \mu_M )$, is given by
\begin{equation}
\label{eq: tau-phi-M}
\begin{split}
&\tau (\lambda) \varphi _M ( \mu_1 ,  \mu_2 ,  \dots ,  \mu_M ) = \chi _M (\mu_1 ,  \mu_2 ,  \dots ,  \mu_M) \varphi _M ( \mu_1 ,  \mu_2 ,  \dots ,  \mu_M )  \\[1ex]
&+ \sum _{i=1}^M \frac{2 \sinh ^2 (2 \lambda)}{\sinh (\lambda + \mu _i) \sinh (\lambda - \mu _i)} 
\frac{\sinh (\xi +\mu _i) \sinh (\xi - \mu _i)}{\sinh (\xi + \lambda)\sinh (\xi -\lambda)} \times \\
& \times %\frac{\sinh (\xi +\mu _i) \sinh (\xi - \mu _i)}{\sinh (\xi + \lambda)\sinh (\xi -\lambda)}
\left( 2 \rho (\mu _i) + \frac{1}{\sinh (\xi + \mu _i)\sinh (\xi -\mu _i)} - \sum _{j\neq i}^M \frac{2}{\sinh (\mu_i + \mu _{j}) \sinh (\mu_i - \mu _{j})} \right) \varphi _M (\lambda, \{ \mu _j \} _{j\neq i} ) , 
\end{split}
\end{equation}
with the eigenvalue
\begin{equation}
\label{eq: chi-M}
\begin{split}
&\chi _M (\mu_1 ,  \mu_2 ,  \dots ,  \mu_M) = \chi _0 (\lambda) - \sum _{i=1}^M  \frac{2 \sinh ^2 (2 \lambda)}{\sinh (\lambda + \mu _i) \sinh (\lambda - \mu _i)} \times \\
&\times \left(2 \rho (\lambda) + \frac{1}{\sinh (\xi + \lambda)\sinh (\xi -\lambda )} - \sum _{j\neq i}^M \frac{1}{\sinh ( \lambda + \mu _{j} ) \sinh (\lambda - \mu _{j} )} \right).   
\end{split}
\end{equation}
The M unwanted terms in \eqref{eq: tau-phi-M} vanish when the Bethe equation are imposed on the parameters $\mu _i$,
\begin{equation}
\label{eq: BE-M}
%\beta _3(\mu _i; \{ \mu_j \} _{j\neq i}) = 
2 \rho (\mu _i) + \frac{1}{\sinh (\xi + \mu _i)\sinh (\xi -\mu _i)}  - \sum _{j\neq i}^M \frac{2}{\sinh (\mu_i + \mu _{j}) \sinh (\mu_i - \mu _{j})}  = 0 ,
\end{equation}
with $i=1,2, \ldots, M$. As expected, the action above of the generating function $\tau (\lambda)$ \eqref{eq: open-tau} is strikingly simple \eqref{eq: tau-phi-M}. This simplicity is due to our definition of the Bethe vector $\varphi _M (\mu _1,  \mu_2 , \ldots,\mu_M)$ \eqref{eq: phi-M} and the corresponding creation operators $\mathcal{C}_K(\mu)$ \eqref{eq: C_K}. In this sense we have completed the implementation of the algebraic Bethe ansatz for the trigonometric Gaudin model, with triangular K-matrix \eqref{eq: K-triangular}, based on the non-unitary classical r-matrix \eqref{eq: rK} and corresponding linear bracket \eqref{eq: linear-rk-br}.

\section{Conclusions \label{sec: Conclu}}
We have derived the generating function of the Gaudin Hamiltonians with boundary terms following the same approach used previously in the rational case, which in turn was based on Sklyanin's method in the periodic case. Our derivation is centered on the quasi-classical expansion of the linear combination of the transfer matrix of the XXZ Heisenberg spin chain and the central element, the so-called Sklyanin determinant. The corresponding Gaudin Hamiltonians with boundary terms are obtained as the residues of the generating function. 

Our next step was the implementation of the algebraic Bethe ansatz for the trigonometric Gaudin model, with triangular reflection matrix \eqref{eq: K-triangular}. To this end we have introduced the non-unitary classical r-matrix \eqref{eq: rK}, which satisfies the generalized classical Yang-Baxter equation \eqref{eq: g-c-YBE}, as well as the modified Lax matrix \eqref{eq: cal-L}. Together they define the linear bracket \eqref{eq: linear-rk-br}, which is obviously anti-symmetric and  it obeys the Jacobi identity. Therefore the entries of the modified Lax matrix generate an infinite dimensional Lie algebra relevant for the open trigonometric Gaudin model. A suitable set of generators \eqref{eq: new-basis-e}-\eqref{eq: new-basis-f} simplifies the commutation relations \eqref{eq: 0-comm}-\eqref{eq: e-f} and therefore facilitates the algebraic Bethe ansatz. Another crucial observation in the implementation of the algebraic Bethe ansatz is the existence of the so-called pseudo-vacuum or the reference state $\Omega _+$ \eqref{eq: Omega+} (see also \eqref{eq: action-Om} and \eqref{eq: tau-Om}). Probably the simplest way to define the relevant Bethe vectors is using the family of the creation operators $\mathcal{C}_K(\mu)$ \eqref{eq: C_K}. These Bethe vectors $\varphi _M (\mu _1,  \mu_2 , \ldots,\mu_M)$ \eqref{eq: phi-M} are symmetric functions of their arguments and they yield strikingly simple off-shell action of the generating function. In this sense, we have fully implemented the algebraic Bethe ansatz, obtaining the spectrum of the generating function and the corresponding Bethe equations. 

It would be of considerable interest to establish a relationship between these Bethe vectors of the trigonometric Gaudin model and solutions to the corresponding generalized Knizhnik-Zamolodchikov equations, analogously as it was done in the case when the boundary matrix is diagonal \cite{Hikami95} but these results will be reported elsewhere.

\bigskip

\noindent
\textbf{Acknowledgments}

\medskip

\noindent
We acknowledge partial financial support by the bilateral scientific cooperation between \break Portugal and Serbia through the project "Quantum Gravity and Quantum Integrable \break Models - 2015-2016", no. 451-03-01765/2014-09/24 supported by the Foundation for Science and Technology (FCT), Portugal, and the Ministry of Education, Science and Technological Development of the Republic of Serbia and by the FCT project PTDC/MAT-GEO/3319/2014. I.S. was supported in part by the Serbian Ministry of Science and Technological Development  under grant number ON 171031.

%\clearpage
%\newpage

%
%
%
\appendix
\section{Basic definitions \label{app: basic-def}}
We consider a spin chain with N sites with spin $s$ representations, i.e. a local $\mathbb{C}^{2s+1}$ space at each site and the operators 
\begin{equation}
S_m^{\alpha} = \mathbbm{1} \otimes \cdots \otimes \underbrace{S^{\alpha}} _m \otimes \cdots \otimes \mathbbm{1},
\end{equation}
with $\alpha = +,-, 3$ and $m= 1, 2 ,\dots , N$. The operators $S^{\alpha}$ with $\alpha = +, - , 3$, act in some (spin $s$) representation space $\mathbb{C}^{2s+1}$ with the commutation relations \cite{KulishResh83,FRT89,Anastasia07,NMIS}
\begin{equation}
\label{eq: cr-spin-s}
[S^3, S^{\pm}] = \pm S^{\pm}, \quad [S^+,S^-] =\frac{\sinh (2\eta S^3)}{\sinh (\eta)} = [2 S^3]_q ,
\end{equation}
with $q = e^{\eta}$ , 
and Casimir operator
\begin{equation}
\label{eq: Casimir}
c_2 = \frac{q+q^{-1}}{2}[S^3]_q ^2 + \frac{1}{2} (S^+S^-+S^-S^+) = \frac{q+q^{-1}}{2}[S^3]_q ^2  
+ \frac{1}{2} [2S^3]_q + S^-S^+ .
\end{equation}
In the space $\mathbb{C}^{2s+1}$ these operators admit the following matrix representation 
\begin{equation}
\label{eq: matrix-rep}
S^{3} = \sum _{i=1}^{2s+1} a_i e_{i\, i} , \quad S^{+} = \sum _{i=1}^{2s+1} b_i e_{i\, i+1} ,
\quad S^{-} = \sum _{i=1}^{2s+1} b_i e_{i+1\, i}  \quad \text{and} \quad c_2  = [s+1]_q \, [s]_q \ \mathbbm{1} ,
\end{equation}
where
\begin{equation}
\label{eq: matrix-elements}
(e_{ij})_{kl} = \delta _{i\,k} \delta _{j\, l}, \quad a_i  = s+1 - i , \quad
b_i = \sqrt{[i]_q \ [2s+1 -i]_q}  \quad \text{and} \quad  [x]_q = \frac{q^x - q^{-x}}{q - q^{-1}}.
\end{equation}
In the particular case of spin $\frac12$ representation, one recovers the Pauli matrices
$$
S^{\alpha} = \frac{1}{2} \sigma ^{\alpha} = \frac{1}{2} \left(\begin{array}{cc}
\delta_{\alpha3} & 2\delta_{\alpha+}  \\
2\delta_{\alpha-} & - \delta _{\alpha 3} \end{array}\right).
$$

\section{Bethe vector $\varphi _3 (\mu _1, \mu _2, \mu _3)$ \label{app: phi3}}

Here we present explicit formulas of the Bethe vector $\varphi _3 (\mu _1, \mu _2, \mu _3)$ in the form of the following symmetric function of its arguments
\begin{equation}
\label{eq: phi-3}
\begin{split}
\varphi _3 (\mu _1, \mu _2, \mu _3) &= f(\mu _1) f(\mu _2) f(\mu _3) \Omega _+ 
+ c_3^{(1)}( \mu _1 ; \mu _2 , \mu _3) f(\mu _2) f(\mu _3) \Omega _+
+ c_3^{(1)}( \mu _2 ; \mu _3 , \mu _1) f(\mu _3) f(\mu _1) \Omega _+ \\
&+ c_3^{(1)}( \mu _3 ; \mu _1 , \mu _2) f(\mu _1) f(\mu _2) \Omega _+
+ c_3^{(2)}(\mu _1 , \mu _2 ; \mu _3) f(\mu _3) \Omega _+
+ c_3^{(2)}(\mu _2 , \mu _3 ; \mu _1) f(\mu _1) \Omega _+ \\
&+ c_3^{(2)}(\mu _3 , \mu _1 ; \mu _2) f(\mu _2) \Omega _+
+ c_3^{(3)}(\mu _1, \mu _2,  \mu _3 ) \Omega _+,
\end{split}
\end{equation}
where the three scalar coefficients above are given by

\begin{align}
\label{eq: c3-1}
c_3^{(1)}( \mu _1 ; \mu _2 , \mu _3) &= \frac{\psi}{\kappa}  \left( 1 + \left( \rho(\mu _1) - \frac{1}{\sinh (\mu_1 - \mu_2) \sinh (\mu_1 + \mu_2)} - \frac{1}{\sinh (\mu_1 - \mu_3) \sinh (\mu_1 + \mu_3)} \right) \times 
\right. \notag \\
& \times \left.  \left( e^{-2\xi} - \cosh( 2 \mu_1 ) \right) \right) , 
\end{align}
\begin{align}
%\\[2ex]
%
%
\label{e: c3-2}
&c_3^{(2)}(\mu _1 , \mu _2 ; \mu _3)  = \frac{\psi ^2}{\kappa ^2} \left( 3 + \left( e^{-2\xi} - \cosh( 2 \mu_1 ) \right)  \left( e^{-2\xi} - \cosh( 2 \mu_2 ) \right) \times 
\right. \notag \\[1ex]
&\times \left( \rho(\mu_1) - \frac{1}{\sinh(\mu_1 - \mu_3) \sinh(\mu_1 +\mu_3)} \right)
\left( \rho(\mu_2)  - \frac{1}{\sinh(\mu_2 - \mu_3) \sinh(\mu_2 +\mu_3)} \right)
\notag \\[1ex]
&+\frac{2 e^{-4\xi} + 2 e^{-2\xi} \left( \cosh(2\mu_1) - 3 \cosh (2\mu_2) \right) - \left( 3 + \cosh (4\mu_1) - 6 \cosh (2\mu_1) \cosh (2\mu_2) \right)}{4 \sinh (\mu_1 - \mu_2) \sinh (\mu_1 + \mu_2)}
\notag \\[1ex]
&\times \left( \rho(\mu_1) - \frac{1}{\sinh(\mu_1 - \mu_3) \sinh(\mu_1 +\mu_3)} \right) 
\notag \\[1ex]
&+ \frac{2 e^{-4\xi} + 2 e^{-2\xi} \left( \cosh(2\mu_2) - 3 \cosh (2\mu_1) \right) - \left( 3 + \cosh (4\mu_2) - 6 \cosh (2\mu_2) \cosh (2\mu_1) \right)}{4 \sinh (\mu_2 - \mu_1) \sinh (\mu_2 + \mu_1)} \notag \\[1ex]
&\left. \times \left( \rho(\mu_2)  - \frac{1}{\sinh(\mu_2 - \mu_3) \sinh(\mu_2 +\mu_3)} \right) \right) , 
\end{align}
\begin{align}
\label{eq: c3-3}
&c _3^{(3)}(\mu _1 , \mu _2 , \mu _3) = \frac{\psi ^3}{\kappa ^3} \left( 15 + \left( e^{-2\xi} - \cosh( 2 \mu_1 ) \right)  \left( e^{-2\xi} - \cosh( 2 \mu_2 ) \right)  \left( e^{-2\xi} - \cosh( 2 \mu_3 ) \right) \rho(\mu_1) \rho(\mu_2) \rho(\mu_3)
\right. \notag \\[1ex]
&+ \left( \left( 5 - 2 \coth(\xi) \left( \frac{\sinh ^2(\mu_1)}{\sinh(\mu_1 - \mu_3) \sinh(\mu_1 +\mu_3)} + \frac{\sinh ^2(\mu_2)}{\sinh(\mu_2 - \mu_3) \sinh(\mu_2 +\mu_3)} \right) \right) 
\left( e^{-2\xi} - \cosh( 2 \mu_1 ) \right)  
\right. \notag \\
&\times
\left( e^{-2\xi} - \cosh( 2 \mu_2 ) \right) -\frac{e^{- \xi}}{\sinh(\xi)} \left( \frac{\sinh (\xi - \mu_1) \sinh (\xi + \mu_1)}{\sinh(\mu_1 - \mu_3) \sinh(\mu_1 +\mu_3)} 
\left( e^{-2\xi} + 1 - 2 \cosh(2\mu_1) \right) \times 
\right. \notag \\
& \left. \left. \times \left( e^{-2\xi} - \cosh( 2 \mu_2 ) \right) + \frac{\sinh (\xi - \mu_2) \sinh (\xi + \mu_2)}{\sinh(\mu_2 - \mu_3) \sinh(\mu_2 +\mu_3)} 
\left( e^{-2\xi} - \cosh( 2 \mu_1 )  \right) \left( e^{-2\xi} + 1 - 2 \cosh(2\mu_2) \right) \right) \right) 
\notag \\
&\times \rho(\mu_1) \rho(\mu_2) 
\notag \\[1ex]
&+ \left( \left( 5 - 2 \coth(\xi) \left( \frac{\sinh ^2(\mu_1)}{\sinh(\mu_1 - \mu_2) \sinh(\mu_1 +\mu_2)} + \frac{\sinh ^2(\mu_3)}{\sinh(\mu_3 - \mu_2) \sinh(\mu_3 +\mu_2)} \right) \right) 
\left( e^{-2\xi} - \cosh( 2 \mu_1 ) \right)  
\right. \notag \\
&\times
\left( e^{-2\xi} - \cosh( 2 \mu_3 ) \right) -\frac{e^{- \xi}}{\sinh(\xi)} \left( \frac{\sinh (\xi - \mu_1) \sinh (\xi + \mu_1)}{\sinh(\mu_1 - \mu_2) \sinh(\mu_1 +\mu_2)} 
\left( e^{-2\xi} + 1 - 2 \cosh(2\mu_1) \right) \times 
\right. \notag \\
& \left. \left. \times \left( e^{-2\xi} - \cosh( 2 \mu_3 ) \right) + \frac{\sinh (\xi - \mu_3) \sinh (\xi + \mu_3)}{\sinh(\mu_3 - \mu_2) \sinh(\mu_3 +\mu_2)} 
\left( e^{-2\xi} - \cosh( 2 \mu_1 )  \right) \left( e^{-2\xi} + 1 - 2 \cosh(2\mu_3) \right) \right) \right) 
\notag \\
&\times
\rho(\mu_1) \rho(\mu_3) 
\notag %\\
\end{align}
\begin{align}
&+ \left( \left( 5 - 2 \coth(\xi) \left( \frac{\sinh ^2(\mu_2)}{\sinh(\mu_2 - \mu_1) \sinh(\mu_2 +\mu_1)} + \frac{\sinh ^2(\mu_3)}{\sinh(\mu_3 - \mu_1) \sinh(\mu_3 +\mu_1)} \right) \right) 
\left( e^{-2\xi} - \cosh( 2 \mu_2 ) \right)  
\right. \notag \\
&\times
\left( e^{-2\xi} - \cosh( 2 \mu_3 ) \right) -\frac{e^{- \xi}}{\sinh(\xi)} \left( \frac{\sinh (\xi - \mu_2) \sinh (\xi + \mu_2)}{\sinh(\mu_2 - \mu_1) \sinh(\mu_2 +\mu_1)} 
\left( e^{-2\xi} + 1 - 2 \cosh(2\mu_2) \right) \times 
\right. \notag \\
& \left. \left. \times \left( e^{-2\xi} - \cosh( 2 \mu_3 ) \right) + \frac{\sinh (\xi - \mu_3) \sinh (\xi + \mu_3)}{\sinh(\mu_3 - \mu_1) \sinh(\mu_3 +\mu_1)} 
\left( e^{-2\xi} - \cosh( 2 \mu_2 )  \right) \left( e^{-2\xi} + 1 - 2 \cosh(2\mu_3) \right) \right) \right) 
\notag \\
&\times
\rho(\mu_2) \rho(\mu_3) 
\notag \\[1ex]
&+ \left( 8 e^{-6\xi} + 4 e^{-4\xi} \left( 4 \cosh (2\mu_1) -5 \left( \cosh (2\mu_2) + \cosh (2\mu_3) \right) \right) + 2 e^{-2\xi} \left( -5 + 7 \cosh (4 \mu_1) + 30 \cosh (2\mu_2) \times \right. \right. \notag \\
& \left. \times \cosh(2\mu_3) - 10 \cosh (2\mu_1) \left( \cosh (2\mu_2) + \cosh (2\mu_3) \right) \right) -3 \cosh (6\mu_1) + 10 \left( 3 + \cosh (4\mu_1) \right) \times
\notag \\
& \Big. \times \left( \cosh (2\mu_2) + \cosh (2\mu_3) \right) - 5 \cosh (2 \mu_1) \left( 5 + 12 \cosh (2\mu_2)  \cosh (2\mu_3)\right) \Big) \times
\notag \\
&\times \frac{\rho(\mu_1)}{16 \sinh(\mu_1 - \mu_2) \sinh(\mu_1 +\mu_2)\sinh(\mu_1 - \mu_3) \sinh(\mu_1 +\mu_3)}
\notag \\[1ex]
&+ \left( 8 e^{-6\xi} + 4 e^{-4\xi} \left( 4 \cosh (2\mu_2) -5 \left( \cosh (2\mu_1) + \cosh (2\mu_3) \right) \right) + 2 e^{-2\xi} \left( -5 + 7 \cosh (4 \mu_2) + 30 \cosh (2\mu_1) \times \right. \right. \notag \\
& \left. \times \cosh(2\mu_3) - 10 \cosh (2\mu_2) \left( \cosh (2\mu_1) + \cosh (2\mu_3) \right) \right) -3 \cosh (6\mu_2) + 10 \left( 3 + \cosh (4\mu_2) \right) \times
\notag \\
& \Big. \times \left( \cosh (2\mu_1) + \cosh (2\mu_3) \right) - 5 \cosh (2 \mu_2) \left( 5 + 12 \cosh (2\mu_1)  \cosh (2\mu_3)\right) \Big) \times
\notag \\
&\times \frac{\rho(\mu_2)}{16 \sinh(\mu_2 - \mu_1) \sinh(\mu_2 +\mu_1)\sinh(\mu_2 - \mu_3) \sinh(\mu_2 +\mu_3)}
\notag \\[1ex]
&+ \left( 8 e^{-6\xi} + 4 e^{-4\xi} \left( 4 \cosh (2\mu_3) -5 \left( \cosh (2\mu_1) + \cosh (2\mu_2) \right) \right) + 2 e^{-2\xi} \left( -5 + 7 \cosh (4 \mu_3) + 30 \cosh (2\mu_1) \times \right. \right. \notag \\
& \left. \times \cosh(2\mu_1) - 10 \cosh (2\mu_3) \left( \cosh (2\mu_1) + \cosh (2\mu_2) \right) \right) -3 \cosh (6\mu_3) + 10 \left( 3 + \cosh (4\mu_3) \right) \times
\notag \\
& \Big. \times \left( \cosh (2\mu_1) + \cosh (2\mu_2) \right) - 5 \cosh (2 \mu_3) \left( 5 + 12 \cosh (2\mu_1)  \cosh (2\mu_2)\right) \Big) \times
\notag \\
&\times \left. \frac{\rho(\mu_3)}{16 \sinh(\mu_3 - \mu_1) \sinh(\mu_3 +\mu_1) \sinh(\mu_3 - \mu_2) \sinh(\mu_3 +\mu_2)} \right)
\end{align}


\clearpage
\newpage


\begin{thebibliography}{55}
\bibitem{Gaudin76} M. ~Gaudin, \textsl{Diagonalisation d'une classe d'hamiltoniens de spin}, J. Physique 37 (1976) 1087--1098.
\bibitem{Gaudin83} M. ~Gaudin, \textsl{La fonction d'onde de {B}ethe}, Masson, Paris, 1983.
\bibitem{Gaudin-English} M. ~Gaudin, \textsl{The Bethe Wavefunction}, Cambridge University Press, 2014.
\bibitem{TakhtajanFaddeev79} L. ~A. ~Takhtajan and L. ~D. ~Faddeev, \textsl{The quantum method for the inverse problem and the $XYZ$ Heisenberg model}, (in Russian) Uspekhi Mat. Nauk \textbf{34} No. 5 (1979) 13--63; translation in Russian Math. Surveys \textbf{34} No.5 (1979) 11--68.
\bibitem{KulishSklyanin82} P. ~P. ~Kulish and E. ~K. ~Sklyanin, \textsl{Quantum spectral transform method. Recent developments}, Lecture Notes in Physics \textbf{151} (1982), 61--119.
\bibitem{Sklyanin89} E. ~K. ~Sklyanin, \textsl{Separation of variables in the {G}audin model}, Zap. Nauchn. Sem. Leningrad. Otdel. Mat. Inst. Steklov. (LOMI) 164 (1987) 151--169; translation in J. Soviet Math. 47 (1989) 2473--2488.
\bibitem{BelavinDrinfeld} A. ~A. ~Belavin and  V. ~G. ~Drinfeld. \textsl{Solutions of the classical Yang-Baxter equation for simple Lie algebras }(in Russian), Funktsional. Anal. i Prilozhen. 16 (1982), no. 3, 1--29; translation in Funct. Anal. Appl. 16 (1982) no. 3, 159-180.
\bibitem{SklyaninTakebe} E.~K. ~Sklyanin and T. ~Takebe, \textsl{Algebraic Bethe ansatz for the XYZ Gaudin model}, Phys. Lett. A 219 (1996) 217-225.
\bibitem{Semenov97} M. ~A. ~Semenov-Tian-Shansky, \textsl{Quantum and classical integrable systems}, in Integrability of Nonlinear Systems, Lecture Notes in Physics Volume 495 (1997) 314-377.
\bibitem{Jurco89} B. ~Jur\v{c}o, \textsl{Classical {Y}ang-{B}axter equations and quantum integrable systems}, J. Math. Phys. Volume 30 (1989) 1289--1293.
\bibitem{Jurco90} B. ~Jur\v{c}o, \textsl{Classical Yang-Baxter equations and quantum integrable systems (Gaudin models)}, in Quantum groups (Clausthal, 1989), Lecture Notes in Phys. Volume 370 (1990) 219--227.
\bibitem{BabujianFlume} H. ~M. ~Babujian and R. ~Flume, \textsl{Off-shell {B}ethe ansatz equation for {G}audin magnets and solutions of {K}nizhnik-{Z}amolodchikov equations}, Mod. Phys. Lett. A 9 (1994) 2029--2039.
\bibitem{FeiginFrenkelReshetikhin} B.~Feigin, E.~Frenkel, and N.~Reshetikhin, \emph{Gaudin model, {B}ethe ansatz and correlation functions at the critical level}, Commun. Math. Phys. 166 (1994) 27--62.
\bibitem{ReshetikhinVarchenko} N.~Reshetikhin and A.~Varchenko, \textsl{Quasiclassical asymptotics of solutions to the {KZ} equations}, in \textrm{Geometry, Topology and Physics}, Conf. Proc. Lecture Notes Geom., pages 293--273. Internat. Press, Cambridge, MA, 1995.
\bibitem{WagnerMacfarlane00} F. ~Wagner and A. ~J. ~Macfarlane,  \textsl{Solvable Gaudin models for higher rank symplectic algebras. Quantum groups and integrable systems (Prague, 2000)} Czechoslovak J. Phys. 50 (2000) 1371--1377.
\bibitem{BrzezinskiMacfarlane94} T. ~Brzezinski and A. ~J. ~Macfarlane,  \textsl{On integrable models related to the osp(1,2) Gaudin algebra}, J. Math. Phys. 35 (1994), no. 7, 3261--3272.
\bibitem{KulishManojlovic01} P. ~P. ~Kulish and N. ~Manojlovi\'c, \emph{Creation operators and {B}ethe vectors of the $osp(1|2)$ {G}audin model}, J. Math. Phys. 42 no. 10 (2001) 4757--4778.
\bibitem{KulishManojlovic03} P. ~P. ~Kulish and N.~Manojlovi\'c, \textsl{Trigonometric $osp(1|2)$ {G}audin model}, J. Math.Phys. 44 no. 2 (2003) 676--700.
\bibitem{LimaUtiel01} A.~Lima-Santos and W.~Utiel, \textsl{Off-shell {B}ethe ansatz equation for $osp(2|1)$ {G}audin magnets}, Nucl. Phys. B  600 (2001) 512--530.
\bibitem{KurakLima04} V.~Kurak and A.~Lima-Santos, \textsl{$sl(2|1)^{(2)}$ Gaudin magnet and its associated Knizhnik-Zamolodchikov equation}, Nuclear Physics B 701 (2004) 497--515.
\bibitem{HikamiKulishWadati92} K. ~Hikami, P. ~P. ~Kulish and M. ~Wadati, \textsl{Integrable Spin Systems with Long-Range Interaction}, Chaos, Solitons \& Fractals  Vol. 2 No. 5 (1992) 543--550.
\bibitem{HikamiKulishWadati92a} K. ~Hikami, P. ~P. ~Kulish and M. ~Wadati, \textsl{Construction of Integrable Spin Systems with Long-Range Interaction}, J. Phys. Soc. Japan Vol. 61 No. 9 (1992) 3071--3076.
\bibitem{Hikami95} K. ~Hikami, \textsl{Gaudin magnet with boundary and generalized Knizhnik-Zamolodchikov equation}, J. Phys. A Math. Gen. \textbf{28} (1995) 4997--5007.
\bibitem{YangZhangSasakic04} W. ~L. Yang,  R. ~Sasaki and Y. ~Z. ~Zhang, \textsl{$\mathbb{Z}_n$ elliptic Gaudin model with open boundaries}, JHEP 09 (2004) 046.
\bibitem{YangZhangSasakic05} W. ~L. Yang,  R. ~Sasaki and Y. ~Z. ~Zhang, \textsl{$A_{n-1}$ Gaudin model with open boundaries}, Nuclear Physics B 729 (2005) 594--610.
\bibitem{YangZhang12} K. ~Hao, W.-L. ~Yang, H. ~Fan, S. ~Y. ~Liu, K. ~Wu, Z. ~Y. ~Yang and Y. ~Z. ~Zhang, \textsl{Determinant representations for scalar products of the XXZ Gaudin model with general boundary terms}, Nuclear Physics \textbf{B 862} (2012) 835--849.
\bibitem{Lima09}A.~Lima-Santos, \textsl{The $sl(2|1)^{(2)}$ Gaudin magnet with diagonal boundary terms}, J. Stat. Mech. (2009) P07025.
\bibitem{Sklyanin86} E. ~K. ~Sklyanin, \textsl{Boundary conditions for integrable equations}, (Russian) Funktsional. Anal. i Prilozhen. 21 (1987) 86--87; translation in Functional Analysis and Its Applications Volume 21, Issue 2 (1987) 164--166.
\bibitem{Sklyanin87} E. ~K. ~Sklyanin, \textsl{Boundary conditions for integrable systems}, in the Proceedings of the VIIIth international congress on mathematical physics (Marseille, 1986), World Sci. Publishing, Singapore, (1987) 402--408.
\bibitem{Sklyanin88} E. ~K. ~Sklyanin, \textsl{Boundary conditions for integrable quantum systems}, J. Phys. A: Math. Gen. \textbf{21} (1988) 2375--2389.
\bibitem{Skrypnyk09} T. ~Skrypnyk, \textsl{Non-skew-symmetric classical r-matrix, algebraic Bethe ansatz, and Bardeen-Cooper-Schrieffer-type integrable systems}, J. Math. Phys. 50 (2009) 033540, 28 pages.
\bibitem{Skrypnyk13}T. ~Skrypnyk,  \textsl{"Z2-graded'' Gaudin models and analytical Bethe ansatz}, Nuclear Physics B 870 (2013), no. 3, 495--529. 
\bibitem{CAMN} N. ~Cirilo Ant\'onio, N. ~Manojlovi\'c and Z. ~Nagy, \textsl{Trigonometric $s\ell(2)$ Gaudin model with boundary terms}, Reviews in Mathematical Physics \textbf{Vol. 25} No. 10 (2013) 1343004 (14 pages).
%
%
%
%
\bibitem{China03} J. ~Cao, H. ~Lin, K. ~Shi and Y. ~Wang, \textsl{Exact solutions and elementary excitations in the XXZ spin chain with unparallel boundary fields}, Nucl. Phys. B 663 (2003)  487--519.
\bibitem{Nepomechie04} R. ~I. ~Nepomechie, \textsl{Bethe ansatz solution of the open XXZ chain with nondiagonal boundary terms}, J. Phys. A 37 (2004), no. 2, 433--440. 
\bibitem{French06} D. ~Arnaudon, A. ~Doikou, L. ~Frappat, E. ~Ragoucy and N. ~Cramp\'e, \textsl{Analytical Bethe ansatz in $gl(N)$ spin chains}, Czechoslovak J. Phys. 56 (2006), no. 2, 141--148.
\bibitem{Martins05} C. ~S. ~Melo,  G. ~A. ~P. ~Ribeiro and M. ~J. ~Martins,  \textsl{Bethe ansatz for the $XXX-S$ chain with non-diagonal open boundaries}, Nuclear Phys. B 711, no. 3 (2005) 565--603.
\bibitem{LucEricRafael07} L. ~Frappat, R. ~I. ~Nepomechie, and E. ~Ragoucy, \textsl{A complete Bethe ansatz solution for the open spin-s XXZ chain with general integrable boundary terms}, Journal of Statistical Mechanics: Theory and Experiment,  0709 (2007)  P09009.
\bibitem{YangWang13} J. ~Cao, W.-L. ~Yang, K. ~Shi and Y. ~Wang, \textsl{Off-diagonal Bethe ansatz solution of the XXX spin chain with arbitrary boundary conditions}, Nuclear Physics B 875 (2013) 152--165.
\bibitem{YangWang13a} J. ~Cao, W.-L. ~Yang, K. ~Shi and Y. ~Wang, \textsl{Off-diagonal Bethe ansatz solutions of the anisotropic spin-1/2 chains with arbitrary boundary fields}, Nuclear Physics B 877 (2013) 152--175.
\bibitem{Eric2013RMP} E. ~Ragoucy, \textsl{Coordinate Bethe ans\"atze for non-diagonal boundaries}, Rev. Math. Phys. 25 (2013), no. 10, 1343007.
\bibitem{Eric13} S. ~Belliard, N. ~Cramp\'e and E. ~Ragoucy, \textsl{Algebraic Bethe ansatz for open XXX model with triangular boundary matrices}, Lett. Math. Phys. 103 No. 5 (2013) 493--506. 
\bibitem{Belliard13} S. ~Belliard and N. ~Cramp\'e \textsl{Heisenberg XXX model with general boundaries: eigenvectors from algebraic Bethe ansatz}, SIGMA Symmetry Integrability Geom. Methods Appl. 9 (2013), Paper 072, 12 pp.
\bibitem{Lima13} R. ~A. ~Pimenta and A. ~Lima-Santos, \textsl{Algebraic Bethe ansatz for the six vertex model with upper triangular K-matrices}, J. Phys. A 46 No. 45 (2013)  455002, 13 pp.
\bibitem{Belliard15} S. ~Belliard, \emph{Modified algebraic Bethe ansatz for XXZ chain on the segment -  I: Triangular cases}, Nuclear Physics \textbf{B 892} (2015) 1--20.
\bibitem{BelliardPimenta15} S. ~Belliard and R. ~A. ~Pimenta \emph{Modified algebraic Bethe ansatz for XXZ chain on the segment - II - general cases}, Nuclear Physics \textbf{B 894} (2015) 527--552.
\bibitem{AvanEtal15} J. ~Avan, S. ~Belliard, N. ~Grosjean and R. ~A. ~Pimenta, \emph{Modified algebraic Bethe ansatz for XXZ chain on the segment - III - Proof}, Nuclear Physics \textbf{B 899} (2015) 229--246.
\bibitem{Nepomechie15} A. ~M. ~Gainutdinov and R. ~I. ~Nepomechie, \textsl{Algebraic Bethe ansatz for the quantum group invariant open XXZ chain at roots of unity}, Nuclear Physics B 909 (2016) 796--839. 
%\bibitem{China13} J. ~Cao, W.- L. ~Yang, K. ~Shi and Y. ~Wang, \textsl{Off-diagonal Bethe ansatz solution of the XXX spin chain with arbitrary boundary conditions}, Nuclear Physics B 875 (2013) 152--165.
\bibitem{China15} X. ~Zhang, Y.-Y. ~Li, J. ~Cao, W.-L. ~Yang, K. ~Shi and Y. ~Wang, \textsl{Bethe states of the XXZ spin-$\textstyle{\frac{1}{2}}$ chain with arbitrary boundary fields} Nuclear Physics B 893 (2015) 70--88. 
\bibitem{CAMS} N. ~Cirilo Ant\'onio, N. ~Manojlovi\'c and I. ~Salom, \textsl{Algebraic Bethe ansatz for the XXX chain with triangular boundaries and Gaudin model}, Nuclear Physics B 889 (2014) 87-108. 
\bibitem{NMIS} N. ~Manojlovi\'c, and I. ~Salom, \emph{Algebraic Bethe ansatz for the XXZ Heisenberg spin chain with triangular boundaries and the corresponding Gaudin model}, Nuclear Physics, \textbf{B 923} (2017) 73-106; \texttt{arXiv:1705.02235}.
\bibitem{Inna14} I. ~Lukyanenko, P. S. ~Isaac, J. ~Links,  \textsl{On the boundaries of quantum integrability for the spin-1/2 Richardson-Gaudin system}, Nuclear Physics \textbf{B 886} (2014) 364-398.
\bibitem{CAMRS} N. ~Cirilo Ant\'onio, N. ~Manojlovi\'c, E. ~Ragoucy and I. ~Salom, \emph{Algebraic Bethe ansatz for the $s\ell(2)$ Gaudin model with boundary}, Nuclear Physics \textbf{B 893} (2015) 305-331.
\bibitem{HaoCaoYang14} K. ~Hao, J. ~Cao, T. ~Yang and W.-L. ~Yang, \textsl{Exact solution of the XXX Gaudin model with the generic open boundaries}, Annals of Physics  354 (2015) 401-408, \texttt{arXiv:1408.3012}.
\bibitem{MNS} N. ~Manojlovi\'c,  Z. ~Nagy and I. ~Salom, \emph{Derivation of the trigonometric Gaudin Hamiltonians}, Proceedings of the 8th Mathematical Physics meeting: Summer School and Conference on Modern Mathematical Physics, MPHYS2014, SFIN \textbf{XXVIII} Series A: Conferences No. A1, ISBN: 978-86-82441-43-4, (2015) 127--135. 
\bibitem{VegaGonzalez}H. ~J. ~de Vega and A. ~Gonz\'alez ~Ruiz, \textsl{Boundary $K$-matrices for the $XYZ$, $XXZ$, $XXX$ spin chains}, J. Phys. A: Math. Gen. \textbf{27} (1994), 6129--6137.
\bibitem{Zamolodchikov94} S. ~Ghoshal and A. ~B. ~Zamolodchikov, \textsl{Boundary S-matrix and boundary state in two-dimensional integrable quantum field theory}, International Journal of Modern Physics A \textbf{09}, 3841 (1994) 3841--3885.
\bibitem{Zamolodchikov94b} S. ~Ghoshal and A. ~B. ~Zamolodchikov, \textsl{Errata: Boundary S-matrix and boundary state in two-dimensional integrable quantum field theory}, International Journal of Modern Physics A \textbf{09}, 4353 (1994) 4353. 
\bibitem{KulishResh83} P. ~P. ~Kulish and N. ~Yu. ~Reshetikhin, \textsl{Quantum linear problem for the sine-Gordon equation and higher representations}, Zap. Nauchn. Sem. Leningrad. Otdel. Mat. Inst. Steklov. (LOMI) Vol. 101 (1981) 101--110, translation in Journal of Soviet Mathematics Volume 23, Issue 4  (1983) Pages 2435-2441.
\bibitem{FRT89} L. ~D. ~Faddeev, N. ~Yu. ~Reshetikhin and L. ~A. ~Takhtajan, \textsl{Quantum groups}, Braid group, knot theory and statistical mechanics, 97--110, Adv. Ser. Math. Phys., 9, World Sci. Publ., Teaneck, NJ, 1989.
\bibitem {Anastasia07} A. ~Doikou, \textsl{A note on the boundary spin s XXZ chain}, Physics Letters A 366 (2007) 556-562.
\end{thebibliography}
\end{document}